# Onset of Vortex Shedding and Hysteresis in Flow over Tandem Sharp-Edged Cylinders of Diverse Cross Sections

M. Kouchakzad[1], A. Sohankar [2,*], M.R. Rastan[3,+]

[1] *Mechanical Engineering Group, Pardis College, Isfahan University of Technology, Isfahan, 84156-83111, Iran*
[2] *Department of Mechanical Engineering, Isfahan University of Technology, Isfahan, 84156-83111, Iran*
[3] *Department of Engineering Mechanics, School of Naval Architecture, Ocean and Civil Engineering, Shanghai Jiao Tong University, 200240, Shanghai, China*

[+] Email: *rastan@sjtu.edu.cn*

**Abstract**
Numerical simulations are conducted to analyze flow characteristics around two tandem sharp-edged cylinders with cross sections of square ($b_1^* = 1$) for the upstream cylinder and rectangle ($b_2^*$) for the downstream cylinder ($b^* = b/$a, where a and b are the sides of cylinders). The study investigates the effects of Reynolds numbers ($Re$ = 30-150), cross-sectional aspect ratios of the downstream cylinder ($b_2^* = 1 - 4$), and scaled gap-spacing between cylinders ($S^*$ = 1-6) on the flow structure, onset of vortex shedding, hysteresis and aerodynamic parameters. The results reveal that increasing $b_2^*$ suppresses the vortex shedding of the upstream cylinder, depending on $S^*$. The suppression is attributed to the interference effect and the adhesion of the shear layers on the downstream cylinder. Three distinct time-mean flow patterns are identified based on the separation and reattachment of shear layers. Flow pattern *I* exhibits parallel flow along the side faces of the upstream cylinder, while the separation bubbles associated with reattachment points are formed in flow pattern *II* on these faces. For pattern *III*, no reattachment point is observed and the separation bubbles cover the upstream cylinder′ side faces. Additionally, two instantaneous flow patterns of extended-body and co-shedding are apperceived within the ranges of examined $Re$ and $S^*$. The behaviors of time-mean and varying forces as well as the vortex shedding frequency are correlated with the flow structures. The onset of vortex shedding and hysteresis dependence are discussed comprehensively. The results show that the critical Reynolds numbers for the onset of vortex shedding decrease from 128±2 to 50±2 with $S^*$ increasing from 1 to 6 ($b_1^* = 1$ and $b_2^* = 4$). The hysteresis limit is found within the range of $3.5 \leq S^* \leq 4.5$ for flow over two tandem cylinders ($b_1^* = 1$ and $b_2^* = 4$) at $Re$ = 150.

## 1. Introduction

Given the importance of the flow over circular and prismatic bluff bodies practically and fundamentally, it is still an active topic although classic. The applications span heat exchangers, coastal structures, and cables at high Reynolds numbers ($Re = U_{in}$a$/v$, where $U_{in}$ and "a" denote freestream velocity and side length) and electronic chips and mini or micro-devices at low-$Re$, to name a few. The complex flow phenomena such as flow separation and reattachment, shear layer and wake instability, and vortex shedding have been investigated widely for 2D and 3D square/circular cylinders in different arrangements (e.g., tandem, side-by-side, staggered) [1-8] and isolated/single ones [9-16]. Although the cross-section significance has been documented already [17], the studies on tandem cylinders mostly focused on twin circular and square ones due to widespread applications and simplicity for flow physics assimilation. The rectangular cross-section and the arrangement of non-homogeneous structures are also omnipresent in human-designed systems; however, they have not yet received the attention that they deserve.

The flow around single or tandem square cylinders is fully attached with no separation at $Re \approx 1$. The steady/laminar flow starts to separate at $Re \approx 5$, generating two symmetric/steady recirculation bubbles behind the cylinders [18]. The bubble size is enlarged monotonically with a $Re$ increment till a critical $Re$ (i.e. $Re_{cr}$) in which the vortex shedding starts. The two streets of oppositely-signed convective spanwise vortices (i.e., vortex shedding) cause oscillatory drag and lift forces on the cylinder. A $Re$ increase augments the amplitudes of the oscillatory force components. Locating an obstacle in the wake of the preliminary prism changes the above scenario entirely. The flow patterns are not simply a function of $Re$, but also gap spacing and the configuration of the downstream prism [19]. For instance, the aerodynamic force unsteadiness and onset of vortex shedding occurs at $Re_{cr} = 45 – 55$ at $S^* = 5$ [18] and $Re_{cr} = 60 – 65$ at $S^* = 4$ [20] for two tandem square cylinders. The corresponding critical $Re$ values for the circular counterpart are $Re_{cr} = 75.5$ and $68$ at $S^* = 2$ and $4$, respectively [1, 21]. It should be noted that the range of the critical Reynolds number or vortex shedding for two tandem rectangular cylinders has not been reported so far. Rastan and Alam [19] thoroughly discussed the flow transition and flow pattern dependence for two tandem circular and square cylinders in $S^* \leq 9$ and $Re < 10^5$. They reported that the flow interference between two cylinders is susceptible to stabilizing the flow with $S^* < S^*_{cr}$, where $S^*_{cr}$ is a critical spacing for the onset of the co-shedding flow regime between the cylinders.

Table 1 lists, the experimental and numerical studies on two tandem square/rectangular cylinders. In this table, the range of Reynolds numbers, cylinder gap-spacing ($S^*$), hysteresis limit, and research findings are presented. This table clarifies that the studies on two tandem sharp-edged cylinders at low Reynolds numbers and the onset of vortex shedding are very limited. Therefore, this study investigates the effect of cylinder gap-spacing and aspect ratio of downstream cylinder on the onset of vortex shedding for two tandem sharp-edged cylinders with different sections.

The three extended/slender-body flow, alternating reattachment flow and co-shedding flow regimes emerge over the tandem cylinders and their occurrence is a function of the $S^*$ and $Re$ values [17-26]. Some researchers reported a fourth regime called bistable flow with a systematic increase in the $S^*$ at high Reynolds numbers of $9.7 \times 10^3$-$6.5 \times 10^4$ [19]. The bistable flow, intermittently swings between two between alternating reattachment and co-shedding flow regimes. For extended/slender-body flow regimes, the upstream-cylinder-generated shear layers either originate at the trailing corners and almost steadily attach to the downstream cylinder at low-$Re$ flows or separate at leading corners and overshoot the downstream cylinder at high-$Re$ flows. The vortices shed only in the wake of the downstream cylinder and a pair of symmetrical bubbles is formed in the gap region. For the reattachment flow, the upstream-cylinder-generated shear layers reattach either side of the downstream cylinder alternatively, while no vortex shedding is seen in the gap. Due to the alternate reattachment, the gap bubbles are asymmetrical. The co-shedding flow regime is characterized by the vortex shedding from both cylinders.

Table 1. A summary of studies on the flow over two tandem square cylinders (except [17, 35], i.e. tandem rectangular cylinders), where Num. and Exp. stand for the numerical and experimental methods. $S^*$: Scaled gap spacing; *OVS*: Onset of vortex shedding; *HL*: Hysteresis limit; *St*: Strouhal number; *GP*: Global parameters.

| Ref. | Method | $Re$ | $S^*$ | HL | Findings |
|---|---|---|---|---|---|
| Sohankar et al. [2] | Exp. | 2033 | 1-6 | - | St, GP |
| Liu et al. [17] | Num. | 150 | 1-8 | - | St, GP |
| Sohankar and Etminan [18] | Num. | $1-200$ | 5 | - | OVS, St, GP |
| Rastan and Alam [19] | Num. | $Re < 10^5$ | 1.5-7 | 4.5-7 | St, GP |
| Sohankar [20] | Num. | $40 - 1 \times 10^3$ | 0.3-12 | 2.75-4.75 | St, GP |
| Liu and Chen [22] | Exp. | $2 \times 10^3 - 1.6 \times 10^4$ | 1.5-9 | 2.5-4.25 | OVS, St, GP |
| Sohankar et al. [23] | Num. | $70 - 150$ | 1-5 | 2-5 | St, GP |
| Sobczyk et al. [24] | Exp. | $4.1 \times 10^3 - 3.294 \times 10^4$ | 2-9 | 3-8.6 | St, GP |
| Sohankar [25] | Num. | $1 \times 10^3 - 1 \times 10^5$ | 1-12 | 2.25-4.25 | St, GP |
| Nikfarjam and Sohankar [26] | Num. | $100 - 150$ | 1-8 | 3.5-4.5 | St, GP |
| Kim et al. [27] | Exp. | $5.3 \times 10^3 - 1.6 \times 10^4$ | 1.5-11 | - | St, GP |
| Yen et al. [28] | Exp. | $300 - 1.1 \times 10^3$ | 1.5-3.5 | - | St, GP |
| Nikfarjam and Sohankar [29] | Num. | $40 - 160$ | 4 | - | GP |
| Xiaiqing et al. [30] | Exp. | $8 \times 10^4$ | 1.25-5 | - | GP |
| Shao et al. [31] | Num. | 100 | 2-6 | - | St, GP |
| Kumar et al. [32] | Num. | 40 | 2-30 | - | St, GP |
| Shang et al. [33] | Num. | $5.3 \times 10^3$ | 4 | - | St GP |
| Shui et al. [34] | Num. | 100 | 1.5-9 | - | GP |
| Islam et al. [35] | Num. | 75-150 | 0.5-10 | - | St, GP |

The flow regime is highly correlated with the engineering parameters such as aerodynamic forces and vortex shedding frequency (i.e. Strouhal number). Therefore, understanding how the flow regime varies with $S^*$ and $Re$ is worthwhile. For example, it has been well-documented that the flow modification from the attachment flow regime to the co-shedding regime is accompanied by an abrupt change in both *St* and fluid force. However, shifting from one flow regime to another is not that straightforward due to hysteresis, where the aerodynamic specifications are highly dependent on the flow history. This phenomenon for the twin tandem cylinders is observed in a specific range of gap spacing (i.e., hysteresis limit or *HL*, see Table 1) [19, 20, 22-26]. In other words, the flow regime in a given $S^*$ and $Re$ is contingent on the past flow state, e.g., varying the gap spacing either in a progressively increasing or decreasing manner. Table 1 illustrates hysteresis pops up in 2.5 ≤ *HL* ≤ 7 if the downstream cylinder is displaced so slowly, regardless of Reynolds numbers [19, 20, 22-26]. Sobczyk et al. [24] observed the range shrinks to *HL* = 3.0 – 6.8 at *Re* = 4100 – 32940 if the downstream cylinder is displaced a bit faster (speed of 0.09 in their study).

Among the studies in Table 1, only Islam et al. [35] and Liu et al. [17] took into consideration the tandem rectangular cylinders; no study has been reported on the sharp-edged tandem cylinders with non-identical cross-section. Islam et al. [35] examined the effect of $b^* = 0.25 – 1.0$, $S^* = 0.5 – 10$, and $Re = 75 – 150$ on the results ($b^* = b/a$, where a and b are the sides of cylinders). Their findings show that the upstream cylinder has a higher drag force compared to that of the downstream cylinder. According to aspect ratio variation, several flow regimes were identified. In unsteady regimes, they found more than one peak in the power spectra of the downstream cylinder. It was reported that the vortex shedding mostly occurs behind the downstream cylinder and the root mean square (RMS) of the lift coefficient for the upstream cylinder is considerably lower than that of the downstream cylinder. Liu et al. [17] also investigated a similar configuration ($b^* = 0.3 – 4$, $S^* = 1 – 8$) at $Re = 150$. The flow patterns were categorized by $S^*$ as a narrow gap ($S^* = 1.0, 2.0$), medium gap ($S^* = 3.0, 4.0$), and wide gap ($S^* = 6.0, 8.0$). It was also found that a larger $b^*$ reduces the fluctuating force acting on the downstream cylinder for narrow gaps, but amplifies for wide gaps. The findings show the narrow gap suppresses the vortex shedding of the upstream cylinder; the aspect ratio has less effect on the fluid force of the two tandem cylinders. Furthermore, the vortex shedding of the downstream cylinder diminishes by increasing $b^*$. They reported that the fluid forces of the downstream cylinder for a wide gap are more dependent on $b^*$, compared to cases with a narrow gap.

Despite some studies on two tandem square [18-20, 22-34] and rectangular [17, 35] cylinders (see Table 1), the above literature review illustrates that the flow around two tandem sharp-edged cylinders with non-similar cross-sections has yet to be investigated. On the other hand, assimilating complex phenomena such as hysteresis [36] and the onset of vortex shedding motivates us to perform low $Re$ simulations. Studying low $Re$ flows is common to dispel unanticipated challenges in the path of developing novel approaches (e.g., [37]) and to generalize the concepts to high Reynolds numbers flows (e.g., [38,39]). Therefore, $Re = 30-150$ and $S^* = 1-6$, $b_1^* = 1$ and $b_2^* = 1 - 4$ are selected to fill the knowledge gap. The specific objectives of this study are (i) illustrating the critical parameters associated with the onset of vortex shedding, (ii) categorizing the flow patterns and correlating them with the engineering parameters, (iii) and eventually, determining the hysteresis limit, and detailing the dependence of the aerodynamic forces on this phenomenon. To this end, the Strouhal number, fluctuating lift and drag coefficients, pressure coefficient, fluctuating normal and shear stresses, and flow patterns are scrutinized in this study.

## 2. Problem definition and formulation

Figure 1 schematically shows the computational domain of the problem under consideration. A rectangle ($b_2^* = 1 - 4$) prism (Cyl2) is arranged in tandem behind a square (a = b or $b_1^* = 1$) one (Cyl1). The dimensions are scaled with "a", and denoted with "*" superscript (e.g., $L_1^* = L_1/a$). Table 2 compares the $Re$, $S^*$, computational domain size, and blockage ratio $\beta = a/h$ between current and previous (two tandem square cylinders) studies. It has been well-documented [9,18] that $\beta \leq 5\%$ is mandatory for keeping the computational results independent of the lateral boundaries' location. $L_1^* = 10$, $L_2^* = 20$, $h^* = 20$, and $S^* = 1 - 6$ were chosen for this study; thus, $\beta = 5\%$.

Table 2. Comparison of the computational domain size (see also Fig. 1) between current and previous (two tandem square cylinders) studies.

| Ref. | Re | S* | $L_1^*$ | $L_2^*$ | h* | β(%) | 2D/3D |
|---|---|---|---|---|---|---|---|
| Sohankar et al. [18] | 1-200 | 5 | 5 | 15 | 20 | 5 | 2D |
| Rastan et al. [19] | $Re < 10^5$ | 1.5-7 | 10 | 25 | 20 | 5 | 2D-3D |
| Sohankar [20] | 40-1000 | 0.3-12 | 7 | 16 | 20 | 5 | 2D-3D |
| Sohankar et al. [23] | 70-150 | 1-5 | 10 | 25 | 20 | 5 | 2D |
| Sohankar [25] | $10^3 - 10^5$ | 1-12 | 8 | 16 | 20 | 5 | 3D |
| Nikfarjam et al. [26] | 100-150 | 1-8 | 8.5 | 16.5 | 20 | 5 | 2D |
| Nikfarjam et al. [29] | 100-160 | 4 | 8.5 | 16.5 | 20 | 5 | 2D |
| Shui et al. [34] | 100 | 1.5-9 | 4 | 21.5 | 18 | 5.5 | 2D |
| Present | 30-150 | 1-6 | 10 | 20 | 20 | 5 | 2D |

The time-dependent continuity and momentum equations in Eqs. 1 – 2 govern the laminar flow motion in the present study. Dealing with the two-dimensional incompressible flow with constant properties in a Cartesian coordinate system, they are expressed as follows:

$$U_{i,i}^* = 0 \tag{1}$$

$$U_{i,\tau}^* + \left(U_i^* U_j^*\right)_{,j} = - P_{,i}^* + Re^{-1} U_{i,jj}^* \qquad i,j = 1,2 \tag{2}$$

$U^*$ and $P^*$ are scaled velocity and pressure. The pressure, time, and velocity have been made dimensionless using $\rho U_{in}^2$ ($\rho$ fluid density), $a/U_{in}$, and $U_{in}$, respectively. The drag coefficient, $C_d$, lift coefficient $C_l$, friction coefficient $C_f$, pressure coefficient $C_p$, and Strouhal number $St$ are defined in Eq. 3.

$$C_d = \frac{F_d}{0.5\rho U_{in}^2 a}, \quad C_l = \frac{F_L}{0.5\rho U_{in}^2 a}, \quad C_f = \frac{\tau_s}{0.5\rho U_{in}^2 a}, \quad C_p = \frac{P-P_\infty}{0.5\rho U_{in}^2}, \quad St = \frac{fa}{U_{in}} \tag{3}$$

The $f, F_d, F_L$, and $P_\infty$ are, respectively, the vortex shedding frequency, drag force, lift force, and freestream pressure. The $\tau_s$ is the wall shear stress. The RMS of lift and drag coefficients are provided in Eq. 4.

$$C_{lrms} = \sqrt{\frac{1}{N}\sum_1^N (C_l(t) - \overline{C}_l)^2} \qquad C_{drms} = \sqrt{\frac{1}{N}\sum_1^N (C_d(t) - \overline{C}_d)^2} \tag{4}$$

The governing equations are numerically solved in a finite volume context using Ansys-Fluent 2019. The diffusion term is discretized by second-order central differencing. The convective term is discretized by Quick's method, which is based on the weighted average of second-order-upwind and central interpolations of the variable. The second-order implicit method is also applied for temporal discretization. The Semi-Implicit Method for Pressure-Linked Equations (SIMPLE) scheme deals with pressure-velocity coupling. The time-marching is performed by steps of 0.02 (see Table 3) and the residual values or convergence criteria for all quantities are set to $10^{-5}$.

A uniform flow is imposed in the x-direction at the inlet. While the zero-gradient velocity sets the outlet of the computational domain. For solid surfaces, the no-slip condition is used. The top and bottom boundaries are modeled as free-slip boundaries, which impose zero normal velocity and zero shear stress. The boundary conditions are summarized as (see also Fig. 1):

Inlet boundary: $u^* = 1, v^* = 0$,  Outlet boundary: $\frac{\partial v^*}{\partial x^*} = \frac{\partial u^*}{\partial x^*} = 0$

Side boundary: $v^* = \frac{\partial u^*}{\partial y^*} = 0$,  Cylinders surface: $u^* = v^* = 0$,

## 3. Validation and grid/time step independence study

The grids with structured tetrahedral cells and non-uniform node spacing were generated. Those next to the solid boundaries (i.e. $\delta^*$ the first grid distance from the cylinder surface) are gradually enlarged with a stretching ratio of less than 1.02 up to 5a in every direction, except 10a, behind the downstream cylinder. The grid size from the middle of the gap spacing shrinks towards the cylinders at a rate of less than 1.05. Further away from this non-uniform grid area, the largest cells (i.e. $\Delta^*$) spatially discretize the computational domain. Figure 2 shows a typical grid generated for the case with $b_1^* = 1$, $b_2^* = 4$, and $S^* = 4$.

If $N_a$ and $N_b$ are respectively the numbers of nodes on the longitudinal sides of Cyl1 and Cyl2, the $N_a = 40$ and $N_b = 60, 70, 80$ for $b_2^* = 2, 3$, and 4 are considered. Furthermore, the number of gap nodes in the x-direction increases as $N_{dis} = 40, 60, 70, 80, 90$, and 100 for $S^* = 1 - 6$ with step one, respectively. To find other specifications for the proper grids, a study was performed for $b_2^* = 4$ by comparing the smallest ($\delta^*$) and largest ($\Delta^*$) grid cell, number of nodes in x ($N_x$) and y ($N_y$) direction and global parameters (i.e., $St$, $\overline{C}_d$, $C_{lrms}$), see Table 3. This table also presents the study of time step influence at $Re = 150$ for a given grid. Except for grids 1 and 2, the results from different grids agree well with each other. The results of grid 4 (71984 nodes) are almost unchanged compared to grids 5 (75856 nodes) and 6 (92700 nodes). Moreover, the hysteresis range was checked with grid numbers 4 and 6. It was observed that both meshes specify the hysteresis limit with the same accuracy. The time step size influence on the results was also examined by studying $\Delta_t^* = 0.01, 0.02$, and 0.03 (see Table 3). The variation of $St$ and $C_{lrms}$ with $\Delta_t^* = 0.02$ and 0.01 are insignificant. However, increasing $\Delta_t^*$ to 0.03 leads to a ~15% discrepancy in $\overline{C}_d$ of Cyl2. Given this examination, $\Delta_t^* = 0.02$ was considered for all simulations.

Table 3. Grid and time step independence tests for the case with $b_1^* = 1$, $b_2^* = 4$, $S^* = 4$, $Re$ =100, 150. $N_x$ and $N_y$ are the node numbers in the x and y direction, and $\Delta_t^*$ shows the scaled time step.

| | $S^* = 4$ | | | | | Cyl1 | | Cyl2 | | |
|---|---|---|---|---|---|---|---|---|---|---|
| Grid | $Re$ | $\delta^*$ | $\Delta^*$ | $N_x*N_y$ | $\Delta_t^*$ | $St$ | $\overline{C}_d$ | $St$ | $\overline{C}_d$ | $C_{lrms}$ |
| 1 | 100 | 0.015 | 0.3 | 320*118 | 0.02 | 0.120 | 1.309 | 0.120 | 1.029 | 0.066 |
| 2 | 100 | 0.01 | 0.25 | 351*150 | 0.02 | 0.119 | 1.290 | 0.118 | 1.005 | 0.052 |
| 3 | 100 | 0.007 | 0.25 | 388*176 | 0.02 | 0.112 | 1.287 | 0.112 | 0.199 | 0.052 |
| 4 | 100 | 0.007 | 0.25 | 409*176 | 0.02 | 0.112 | 1.280 | 0.112 | 0.199 | 0.051 |
| 5 | 100 | 0.005 | 0.2 | 431*176 | 0.02 | 0.112 | 1.280 | 0.112 | 0.199 | 0.051 |
| 6 | 100 | 0.005 | 0.18 | 450*206 | 0.02 | 0.112 | 1.280 | 0.112 | 0.199 | 0.051 |
| 5 | 150 | 0.005 | 0.2 | 431*176 | 0.02 | 0.120 | 1.428 | 0.120 | -0.028 | 0.100 |
| 6 | 150 | 0.005 | 0.18 | 450*206 | 0.02 | 0.120 | 1.427 | 0.120 | -0.028 | 0.100 |
| 4 | 150 | 0.007 | 0.25 | 409*176 | 0.01 | 0.120 | 1.428 | 0.120 | -0.028 | 0.100 |
| 4 | 150 | 0.007 | 0.25 | 409*176 | 0.02 | 0.120 | 1.428 | 0.120 | -0.028 | 0.100 |
| 4 | 150 | 0.007 | 0.25 | 409*176 | 0.03 | 0.121 | 1.410 | 0.121 | -0.024 | 0.101 |

For further grid inspection, the surface pressure coefficient and the mean velocity on the centerline ($y = 0$) are compared at $Re$ =150 for grids in Table 3. Although a satisfactory agreement between $\overline{C}_p$ profiles is seen for Cyl1 in Fig. 3a, grid 1 shows a maximum 61% discrepancy for Cyl2, compared with those of grids 3-6 in Fig. 3b. At x > 15 the $u_{avr}$ on the center line ($y = 0$) for grids 1 and 2 compared to those of grids 3-6 have maximum differences about 50% and 17%, respectively. While it is the same for x < 15 (Fig. 3 c). In other words, the local characteristics (i.e. $\overline{C}_p$) and mean parameters (i.e. $u_{avr}$) for grid 3-6 are independent of the grid number. Therefore, grid 4 was chosen to keep the accuracy and optimize the computation time. Grid 4 specifications (e.g. expansion ratio and the smallest cell size) are used to generate the grids of other $b_2^*$ values.

Due to the lack of appropriate data for verification, Fig. 3 compares the $\overline{C}_d$ of tandem square cylinders at different $S^*$ and $Re = 150$ with the previous attempts [17, 20, 26] for validation purposes. Aerodynamic parameters such as drag coefficient are bifurcated in the cylinder gap-spacing $S^* = 2-5$, which shows the effect of hysteresis on the flow (Fig. 3 (d, e)). The results have good compliance (mostly less than 3.5%) with previous studies in the hysteresis limit (mode1 & mode2).

Table 4 compares (a) the results of flow over a single square cylinder and (b) two tandem square cylinders with $S^* = 4$ at $Re = 100$ & 150 with available data in the literature. This comparison shows that the global results for a single cylinder are in good agreement (mainly less than 8% for $C_{lrms}$ and 5% for other parameters) with those of previous works. Moreover, flow over two tandem square cylinders (Re = 100- Table 4 (b)) is also compared with the reported results of flow over tandem square cylinders [20, 26]. The results are in good agreement (the percentage deviation less than 3%) with the previous studies.

Table 4. Comparison of our results with previous publications for (a) a single square cylinder and (b) two tandem square cylinders with $b_1^* = b_2^* = 1$ ($Re$ =100, 150).

(a)

| Ref. | Re | $\overline{C}_d$ (Max difference) | $St$ (Max difference) | $C_{drms}$(Max difference) | $C_{lrms}$(Max difference) |
|---|---|---|---|---|---|
| Mashhadi et al [14] | 100 | 1.503 | 0.148 | 0.008 (2.1%) | 0.184 |
| Sohankar et al [18] | 100 | 1.533 (2.6%) | 0.149 | 0.007 | 0.204 (7.9%) |
| Sharma et al. [40] | 100 | 1.495 | 0.148 | 0.006 | 0.192 (1.5%) |
| Sen et al [41] | 100 | 1.530 | 0.145 (1.4%) | - | 0.193 |
| Present | 100 | 1.493 | 0.147 | 0.0054 | 0.189 |
| Aboueian et al. [3] | 150 | 1.453 | 0.158 | 0.0156 | 0.264 (8.3%) |
| Mashhadi et al [14] | 150 | 1.466 | 0.162 | 0.017 (4.2%) | 0.269 |
| Sharma et al. [40] | 150 | 1.465 | 0.158 | 0.016 | 0.291 |
| Zheng and Alam [42] | 150 | 1.483 | 0.158 | 0.0163 | 0.277 |
| Alam et al [43] | 150 | 1.492 (3.1%) | 0.154 (3.1%) | 0.0155 (5%) | 0.271 |
| Present | 150 | 1.447 | 0.159 | 0.0163 | 0.288 |

(b)

| | $S^* = 4$ | | Cyl1($b_1^* = 1$) | Cyl2 ($b_2^* = 1$) |
|---|---|---|---|---|
| Ref. | Re | $St$ | $\overline{C}_d$ | $\overline{C}_d$ |
| Sohankar et al. [20] | 100 | 0.139 (2.2%) | 1.472 (0.9%) | 1.178 (2.7%) |
| Nikfarjam et al. [26] | 100 | 0.135 (0.8%) | 1.434 (2.3%) | 1.138 (0.8%) |
| Present | 100 | 0.136 | 1.458 | 1.147 |

## 4. Results and discussion

Numerical simulations are performed at $Re$ = 30-150 and $S^* = 1$-6 to study gap and wake flows of square ($b_1^* = 1$) and rectangle ($b_2^* = 1 - 4$) prims in a tandem arrangement. The critical Reynolds number ($Re_{cr}$) associated with the transition from steady to unsteady flow is determined first. Then, the flow patterns are categorized and the engineering parameters, such as aerodynamic force and shedding frequency, are discussed thoroughly. Finally, the hysteresis at $Re$ = 150 is scrutinized.

### 4-1 Onset of vortex shedding ($Re_{cr}$)

#### 4-1-1 $Re_{cr}$ for various $S^*$ ($b_1^* = 1, b_2^* = 4$)

The flow is steady when the $Re$ is less than a critical one. As such, there are no fluctuations in lift and drag signals. Figure 4 (a, b) shows the lift and drag coefficients at $Re$ = 70, $S^* = 5$, and, $b_2^* = 4$, where no fluctuation is observed. The flow is thus steady at $Re \leq 70$ for $S^* = 5$. Oscillating the lift and drag coefficients for both cylinders at $Re$ = 75 is periodic (Fig. 4 (b, c)), which is a signature of unsteady flow and shedding the vortices. As a result, the onset of vortex shedding for $S^* = 5$ transpires at 70 < $Re$ < 75.

With a similar approach to determine the $Re_{cr}$ range, Fig. 4 depicts the vortex shedding inception threshold for various $S^*$ or $b_2^*$ values. Given the uncertainty of this method (e, g. 70 < $Re_{cr}$ < 75), the mid-value has been assumed to serve as $Re_{cr}$, e.g., $Re_{cr} = 73 \pm 2$ for $S^* = 5$, $b_1^* = 1$, $b_2^* = 4$ (see Fig. 4). Figure 5 (a) illustrates the $Re_{cr}$ ($S^* = 1 - 6$, $b_1^* = 1$, $b_2^* = 4$). But the vortices do not shed necessarily from the upstream cylinder at $Re$ between 30 to 150 and $S^* = 1 - 4$. While it is the case for $S^* = 5, 6$ and the $Re_{cr}$ is the same for both cylinders. Nonetheless, with a focus on the shedding in the wake of the downstream cylinder, the $Re_{cr} = 128 \pm 2$, 119±2, 113±2, 93±2, 73±2, 50±2 for $S^* = 1, 2, 3, 4, 5, 6$, respectively.

The $Re_{cr}$ at $S^* = 6$ approaches that of a single square cylinder with $Re_{cr} = 51.2$ [9]. This observation is ascribed to the mitigation of the interference effect, where the cylinder in the wake of another at higher $S^*$ behaves as an isolated obstacle. The results in Fig. 5 (a, b) are fitted with a curve, and corresponding relations are presented in Eqs. 6 and 7. Eq. 6 is valid for the downstream cylinder, irrespective of $S^*$, but only for the upstream cylinder at $S^* = 5, 6$. Eq. 7 represents the $Re_{cr}$ for $b_2^* = 1 - 4$. The deviation between the data and curve fits at Eqs. 6 and 7 is less than 3% and 0.5%, respectively.

$$Re_{cr} = 131.4 - 1.185 S^* - 2.0714 \, S^{*2} \tag{6}$$

$$Re_{cr} = -191 + 385 b_2^* - 128.5 \, b_2^{*2} + 12.5 \, b_2^{*3} \tag{7}$$

**4-1-2 $Re_{cr}$ for various $b_2^*$ ($S^* = 4, b_1^* = 1$)**

This section reports the $Re_{cr}$ values for $S^* = 4$ and various $b_2^*$. Fig. 5 (b) shows $Re_{cr} = 78 \pm 2$ for both cylinders when $b_1^* = b_2^* = 1$. This $Re_{cr}$ concords well with 80±10 from Ref. [19]. The vortices do not shed from the upstream cylinder but it occurs for the downstream cylinder with $b_2^*$ of $2 - 4$ ($S^* = 4$, $b_1^* = 1$). Thus, an increase of $b_2^*$ ceases the upstream cylinder vortex shedding. The critical Reynolds numbers for the downstream cylinder are $Re_{cr} = 165\pm2$, 145±2, 93±2 for $b_2^* = 2, 3, 4$, respectively. This figure also reveals that the $Re_{cr}$ increases with the $b_2^*$ increment from 1 to 2, and then decreases. Therefore, at a fixed $S^*$, the onset of the vortex shedding is postponed first by increasing $b_2^*$, and then approaches that of equal tandem cylinders. In other words, adding a cylinder in the wake of another controls the upstream cylinder wake and suppresses or delays the vortex shedding. It is worth noting that 140 unsteady simulations for various $Re$, $S^*$, and $b_2^*$ were performed to find the $Re_{cr}$.

**4-2 Flow patterns**

**4-2-1 Instantaneous flow pattern**

The identification of vortex can be carried out by analyzing the streamlines, contours of vorticity, and Q (Figs. 6 and 7). The Q criterion defines the vortex core as a continuous region with a positive second invariant ($\nabla u$) [44, 45]. It should be emphasized that the Q criterion and vorticity provide relatively similar vortex cores for 2D simulations; thus, the streamlines and contours of vorticity are mostly discussed. Figure 6 shows the streamlines and vorticity contours ($\omega_z^*$, scaled by "a" and $U_{in}$) at an instant in which the lift coefficient is maximum. Beyond the zones in which the flow is steady or $Re > Re_{cr}$, two distinct flow patterns of the extended-body (*EBF*) and co-shedding (*CSF*) are found as shown in Fig. 6. For the former (*EBF*), the upstream-cylinder-generated shear layers steadily glide on the downstream cylinder, separate from its trailing edges, roll up in the wake alternatively and form a single vortex street behind the downstream cylinder, see Fig. 6 (a-d, i, j). For the latter (*CSF*), the upstream-cylinder-generated shear layers transform into vortices and shed alternately in the gap (Fig. 6 (e, f, g)). Vortex shedding occurs behind both cylinders in this case.

The steady flow was observed at $S^*$ =1-6 and $Re < Re_{cr}$, e.g. see Fig. 6 (h). The extended-body flow occurs for $S^* < 5$ and $75 \leq Re \leq 150$ at $b_2^* = 4$. This flow pattern is seen when the gap spacing is less than a critical value, i.e., $S^* < S_{cr}^*$. A pair of symmetric vortices are formed between two cylinders for $S^* \leq 4$ ($b_2^* = 4$) and no vortex sheds from the upstream cylinder. While the vortex shedding is observed at $Re > Re_{cr}$ behind the downstream cylinder. The reattachment flow pattern is not observed at $S^* = 5, 6$ for $Re > Re_{cr}$. The critical gap spacing is between 4 and 5 for $b_2^* = 4$. Beyond that, the co-shedding flow is observed at $Re$ = 75-150 for $S^* \geq 5$ and $Re$ = 55-150 for $S^* \geq 6$.

The previous studies for tandem square cylinders ($b_1^* = b_2^* = 1$) widely reported the alternating reattachment flow at a certain range of $S^*$ [18, 19, 22]. This flow pattern is not observed in this study, and the flow pattern transits from extended-body flow to co-shedding directly (Fig. 6). The rectangle cross-section ($b_2^* = 4$) of the downstream cylinder likely modulates the alternation of the upstream cylinder-generated shear layers and ceases the swinging motion of reattachment points on the lateral cylinder faces. Because of the same reason, the two gap bubbles are steadily identical. Whereas, the asymmetry of these bubbles has been reported [19] for the reattachment flows for the twin tandem cylinders.

For a given gap spacing of $S^* = 4$, the flow structure dependence on $b_2^* = 1 - 4$ is discussed here. As mentioned, the flow is steady at $Re < Re_{cr}$, e.g. for $b_2^* = 1 - 4$ ($S^* = 4$). The results show that the co-shedding flow emerges for $b_1^* = b_2^* = 1$ at $Re_{cr} \leq Re \leq 150$ (Fig. 6g), while the extended-body flow occurs for $b_1^* = 1$, $b_2^* = 3 - 4$ at $Re_{cr} \leq Re \leq 150$ (Fig. 6(i, j)). Interestingly, the flow over both cylinders remains steady at $b_2^* = 2$ up to $Re$ = 160 (see Fig. 6h), i.e., the vortex shedding associated with the extended-body flow is postponed up to $Re_{cr} \geq 165\pm2$. The alternating reattachment flow was not observed for any cases ($S^* = 4$, $b_1^* = 1$, $b_2^* = 1 - 4$, $Re \leq 150$).

The values of vorticity at the center of vortices are also shown in Fig. 6 (g-j) for $Re$ = 150, which decreases along the streamwise. Moreover, Fig. 6 (g, h) illustrates that the flow is unsteady at $b_2^* = 1$ and steady at $b_2^* = 2$. Whereas, the gap flow is quasi-steady at $b^* = 3$ and 4. The result shows that the co-shedding flow occurs for $Re > 75$ at $b_1^* = b_2^* = 1$ (e.g. see Fig. 6 (g)), but not anymore for $b_2^* > 1$ at any given $Re$ up to 150 ($Re_{cr} = 165\pm2$). It means that increasing $b_2^*$ suppresses the vortex shedding from both cylinders at $b_2^* = 2$ (e.g. at $Re$ = 150, Fig. 6 (h)), and then the extended-body flow pattern pops up for $b_2^* = 3, 4$ at $Re \geq 140, 90$, respectively.

Figure 7 (a,b) shows the maximum vorticity magnitude $|\omega_z^*|_{max}$ of the detached vortices (identified by Q-criterion in Fig. 7 (c,d)) behind the downstream (Cyl2) and upstream (Cly1) cylinders, available for both of them in the co-shedding flow and only for the downstream one in the extended-body flow. For the extended-body flow ($S^* \leq 4$), the $|\omega_z^*|_{max}$ values are almost constant irrespective of $S^*$ but they sharply increase at $4 < S^* \leq 5$ by transforming the flow patterns to co-shedding flow (Fig. 7 (a)). For $S^* \geq 5$, the $|\omega_z^*|_{max}$ values slightly decrease for both cylinders (Fig. 7 (a)). A sudden drop in the $|\omega_z^*|_{max}$ of the downstream cylinder is observed in Fig. 7(b) with increasing $b_2^*$ from 1 to 2. This variation is also attributed to the flow pattern change, but in this case from co-shedding to steady flow, see also Fig. 6 (g, h). Then the $|\omega_z^*|_{max}$ has a small variation with increasing $b_2^*$ from 2 to 4, where the extended-body flow occurs (see Fig. 6 (i, j)).

The instantaneous results are reported in Fig. 6 only for $Re=150$. To find mean/instantaneous flow patterns and global parameters for various $Re$, $S^*$, and $b_2^*$, many unsteady simulations are carried out for $Re = 30$-$150$, $S^* = 1 - 6$, and $b_2^* = 1 - 4$. All simulations performed for various $Re$ and $S^*$ ($b_2^* = 4$, $b_1^* = 1$) and $Re$ and $b_2^*$ ($S^* = 4$, $b_1^* = 1$) are shown in Figs. 8 a and b, respectively. In this $Re - S^*$ or $Re - b_2^*$ plane, the squares specify simulated cases and the gray and yellow regions illustrate the extended-body (*EBF*) and co-shedding (*CSF*) flow patterns, respectively. The green line in this figure shows the positions of $Re_{cr}$ for each $S^*$ or $b_2^*$, where this boundary separates the steady and unsteady (including both *EBF* and *CSF* regions) flow zones. These flow maps show a comprehensive overview of all steady/unsteady flow patterns that occur for each $Re$, $S^*$ and $b_2^*$.

### 4-2-2 Mean flow patterns

Figures 9 and 10 show the time-mean flows at $b_1^* = 1$ and $b_2^* = 4$ for various $Re$ and $S^*$. Due to the elongation of the downstream cylinder ($b_2^* = 4$), the mean streamlines around the downstream cylinder are almost the same regardless of $Re$ and $S^*$ values. Therefore, the mean flow patterns are categorized according to the separation and reconnection on the upstream cylinder. It should be noted that the instantaneous (Fig. 8 (a)) and mean flow patterns (Fig. 9 (a)) regions around the critical Reynolds numbers overlap with each other. The $Re_{cr}$ lines in this figure separate the steady and unsteady flow zones. The blue squares in Fig. 9 (a) specify the simulated cases and the three time-mean flow patterns are named *I*, *II*, *III*. The variation of time-mean flow characteristics such as separation and reattachment points with $Re$, $b_2^*$, and $S^*$ are investigated in this investigation. The separation (*SP*), reattachment (*RP*), and saddle (*SD*) points are marked in Fig. 9 (b-d). Some parameters are defined in Fig. 10(a,d) for quantifying the time-mean flow features, including the gap and wake bubble lengths $L_r^*$ and widths $W_r^*$, the side bubble length $L_s^*$, the separation point location $L_d^*$, and the side bubble width $W_s^*$. It should be noted that all lengths are scaled with "a", see Fig. 1.

Mean flow pattern *I* show no reattachment or separation bubble on the side faces of both cylinders and the streamlines are parallel to them. The mean flow is separated from the cylinders' rear corners (*RC*) and a pair of recirculation bubbles form in both gap and wake regions (Fig. 9 (b)). For example, this flow pattern is observed in the range of $Re = 30$-$95$ at $S^* = 1 - 4$ ($b_2^* = 4$), see Table 5 and Fig. 10 (a) for more details. It should be emphasized that Table 5 lists the data from Fig. 10 for $b_1^* = 1$, $b_2^* = 4$, and $S^* = 4$ in $Re = 30$-$150$.

Flow separates from the leading corners (*FC*) of the upstream cylinder in pattern *II*, reattaches on the side faces, and forms a pair of recirculation bubbles on them (Fig. 9 (c)). But, similarly to flow pattern *I*, there is no reattachment or separation bubble on the counterpart faces of the downstream cylinder and the flow separates from the trailing corners. For instance, this flow pattern is formed in *Re* = 100-105 and $S^* = 4$ ($b_2^* = 4$). Table 6 and Fig. 10 (b, c) provide additional details. The results of pattern *II* show the reattachment points move toward the rear corners of the upstream cylinder and the bubble size is enlarged as *Re* increases (e.g. see Fig. 10(b, c)).

Flow also separates in pattern *III* from the leading corners of the upstream cylinder but without any reattachment (Fig. 9 (d)). The flow over the downstream cylinder is the same as the two other flow patterns of *I* and *II* ($b_2^* = 4$). The side separation bubbles almost cover the whole side face in flow pattern *III* and due to the collision of the separation bubble on the side surface of the cylinder with the bubble in the gap-spacing, a saddle point (*SD*) emerges adjacent to the rear corners. On the other hand, The *SD* point is a result of the confrontation of the downward motion of the shear layer and the upward motion of the gap flow (Figs. 9 (d), 10 (d)). For example, flow pattern *III* is observed at *Re* = 110-150 ($S^* = 5$, $b_2^* = 4$). Table 6 provides the *Re* and $S^*$ ranges ($b_2^* = 4$) for the three time-mean flow regimes, where flow pattern *III* is absent at $S^* = 5, 6$.

Besides, Fig. 10 illustrates that a pair of symmetrical recirculation bubbles is formed behind the rectangular cylinder in the mean flow ($b_2^* = 4$ and $S^* = 5$). Based on Fig. 10, the size of this recirculation bubble increases up to *Re* = 150 with increasing *Re* and then decreases (see Fig. 11).

The previous studies for flow over a single square cylinder reported that flow separation takes place from the rear and front edges of the cylinder at *Re* = 100 and 150, respectively [9, 14]. Interestingly, this behavior is not similar to that occurs for the upstream cylinder in two tandem cylinders. Based on the mean and instantaneous streamlines consideration in the present work for $S^* = 1\text{-}4$ and *Re* = 100 ($b_1^* = 1$, $b_2^* = 4$), the flow separation occurs from the front edge and the separation bubbles form on the side surfaces of the upstream cylinder (flow pattern *II*, see Fig. 9). At *Re* = 100, for $S^* = 5\text{-}6$, the flow separates from the rear edge without the formation of separation bubbles on the side surfaces of the upstream cylinder (flow pattern *I*). For *Re* > 110 and $S^* = 5\text{-}6$, the flow pattern changes from *I* to *II* (Fig. 9 (a)), where separation occurs from the front edge and the separation bubbles form on the side surfaces of the upstream cylinder (see Fig. 9 for more detail about the position of separation points for various *Re* and $S^*$).

Table 5. The parameters of mean flow patterns for various *Re* ($b_1^* = 1$, $b_2^* = 4$, $S^* = 4$), see also Figs. 9 and 10 (Cyl1: upstream cylinder, Cyl2: downstream cylinder, RC: rear corner, FC: front corner, SP: separation point).

| *Re* | SP (Cyl1) | $W_s^*$ (Cyl1) | $L_s^*$ (Cyl1) | $L_d^*$ (Cyl1) | SP (Cyl2) | Flow pattern |
|---|---|---|---|---|---|---|
| 30-95 | RC | - | - | - | RC | *I* |
| 95 | RC | - | - | - | RC | *I* |
| 105 | RC & FC | 0.03 | 0.53 | 0.25 | RC | *II* |
| 110 | FC | 0.05 | 0.84 | 0.20 | RC | *III* |
| 120 | FC | 0.07 | 0.55 | 0.12 | RC | *III* |
| 130 | FC | 0.09 | 0.49 | 0.10 | RC | *III* |
| 140 | FC | 0.10 | 0.29 | 0.07 | RC | *III* |
| 150 | FC | 0.12 | 0.11 | 0.05 | RC | *III* |

Table 6. The time-mean flow pattern dependence on the $Re$ and $s^*$ ($b_2^* = 4$).

| Pattern | (I) | (II) | (III) |
|---|---|---|---|
| $S^*$ | | $Re$ | |
| 1 | 30-95 | 100-115 | 120-150 |
| 2 | 30-95 | 100-115 | 120-150 |
| 3 | 30-95 | 100-105 | 110-150 |
| 4 | 30-95 | 100-105 | 110-150 |
| 5 | 30-110 | 120-150 | - |
| 6 | 30-115 | 120-150 | - |

Figures 10 and 11 provide further insights into the time-mean wake flow. The findings illustrate that the $L_r^*$ and $W_r^*$ change with $Re$ and $S^*$. Fig. 11(a) shows that $L_r^* = S^*$ for the upstream cylinder when $S^* \leq 4$, regardless of $Re$, with an exception at $Re = 30$ and $S^* = 4$. On the other hand, shedding the vortices in the gap in the instantaneous sense (e.g. for $S^* = 5,6$) comes with not filling the gap in the time-mean sense (Fig. 10 (b, c, d)). For $S^* \leq 4$ as the $Re$ increases, the $W_r^*$ for the upstream cylinder increases with a relatively mild slope (Fig. 11 (c)). At $S^* = 5$, the $W_r^*$ increases with $Re$ at first, then suddenly decreases, and afterward varies slightly (Fig. 11 (c)). The variations are similar for $W_r^*$ and $L_r^*$ at $S^* = 5$, as the flow pattern changes from extended-body to co-shedding ($Re = 75$-$70$). At $S^* = 5$ with gap vortex shedding, the variations of $W_r^*$ and $L_r^*$ with $Re$ are not much (Figs. 11 (a, c)). For the downstream cylinder at $S^* \leq 4$, where the flow is extended-body one, the $L_r^*$ increases up to $Re = 120$ and then decreases (Fig. 10 (b)). The reason for this increase and decrease is related to the base pressure variation of the downstream cylinder. In co-shedding flow for $S^* > 4$, the $L_r^*$ increases with $Re$ (Fig. 11 (b)). The $W_r^*$ for this cylinder mostly increases with $Re$ (Fig. 11 (d)).

The time-mean flow pattern is also understood with the observation of the $L_s^*$ variation (Fig. 10 (d) and Table 5). The separation bubble on the side faces of the upstream cylinder is formed for $S^* = 1$-$4$ in the range of $100 \leq Re \leq 150$ and for $S^* = 5, 6$ in $Re \geq 120$. In the extended-body flow pattern ($S^* \leq 4$), the $L_s^*$ increases with $Re$ until $Re = 120$ (affiliated to $S^*$). However, for $Re \geq 120$, it reduces (see e.g. for $S^* = 4$ in Table 5). In the co-shedding flow ($S^* \geq 5$), the $L_s^*$ elongates up to $Re = 150$. The $L_s^*$ in $S^* \geq 5$ with unsteady flow has no variation with increasing $S^*$ and is fixed. Generally, the $L_d^*$ shrinks with increasing $Re$ in both extended-body and co-shedding flows. Besides, the $L_d^*$ is present for mean flow patterns $II$ and $III$.

The mean flow patterns were also investigated for a constant gap spacing ($S^* = 4$) and various $b_2^* = 1 - 4$ ($b_1^* = 1$) and $Re = 30$-$150$. Similar to Figs. 9 and 10, three mean flow patterns $I$, $II$, and $III$ were observed for various $Re$ and $b_2^*$. For example, flow patterns $I$ and $II$ occur for $Re < 100$ and $Re \geq 100$, respectively, at $b_2^* = 1$. They were $I$ and $III$ for $Re \leq 100$ and $Re > 100$, respectively, at $b_2^* = 3$. For $b_2^* = 2$, all three mean flow patterns were observed. It should be noted the results for $b_2^* = 4$ are similar to those reported in Figs. 9 and 10 for $S^* = 4$.

### 4-2-3 Fluctuating stresses

Fluctuating stresses are informative for assimilating the flow behavior. The Reynolds decomposition of velocity gives the time-mean and fluctuating components. Here, similar to turbulent flows, the fluctuating normal stress components ($\overline{v^*v^*}$ and $\overline{u^*u^*}$), the shear stress $\overline{u^*v^*}$, and the kinetic energy $KE^* = 0.5(\overline{u^*u^*} + \overline{v^*v^*})$ are defined. Figure 12 (a-f) shows the $\overline{u^*u^*}$ and $\overline{v^*v^*}$ contours for $S^* = 1, 4$ and 6 at $Re = 150$. Since those for $S^* = 2, 3$ and $S^* = 5$ respectively give no new insight compared with those of $S^* = 1$ and $S^* = 6$, they have been skipped in this figure. Fig. 12 (g) represents the $(\overline{u^*u^*})_{max}$ and $(\overline{v^*v^*})_{max}$ values.

The $\overline{u^*u^*}$ and $\overline{v^*v^*}$ values are close to null in the gap region of cases with $S^* \leq 4$, where the vortices do not shed. They come into being at $S^* > 4$ with the co-shedding flow. The two peaks for $\overline{u^*u^*}$ and one for $\overline{v^*v^*}$ are conspicuous in the wake of the downstream cylinder, regardless of $S^*$. Accordingly, Fig. 12(g) shows that the $(\overline{u^*u^*})_{max}$ and $(\overline{v^*v^*})_{max}$ are relatively small and fixed in the extended-body flow ($S^* \leq 4$) and increase or emerge by turning the flow pattern into co-shedding ($S^* > 4$).

The streamwise distances between the cylinders and the $(\overline{u^*u^*})$ and $(\overline{v^*v^*})$ peaks are denoted as the vortex formation length $L_f^*$ and cross-stream oscillation length $L_e^*$, see the definitions in Fig. 12 (a, b, e, f)) for both cylinders. The $L_f^*$ has an inverse relationship with the mean drag coefficients as well as the RMS of drag and lift for a single cylinder [46, 47]. This inverse relationship is the case between the $L_e^*$ and vortex shedding frequency [14]. More details are provided in sections 4-3-3 and 4-3-4. Fig. 12 (h) portrays the $L_f^*$ and $L_e^*$ variations versus $S^*$. The $L_f^*$ is present for the upstream cylinder only if $S^* \geq 5$ (i.e. co-shedding flow). But, in the wake of the downstream cylinder, the larger $S^*$ provides the longer $L_f^*$ in the extended-body flow. The trend is reversed when the flow changes to the co-shedding (i.e. $S^* \geq 5$). The $L_e^*$ steadily increases with $S^*$ although it experiences a mild variation at $4 \leq S^* \leq 5$, where the flow pattern changes. The $L_e^*$ and $L_f^*$ for the downstream cylinder are much larger than those for the upstream cylinder in the co-shedding flow (Fig. 12 (h)).

Figure 13 (a-d) displays the shear stress contours $\overline{u^*v^*}$ at $S^* = 1\text{-}6$ and $Re = 150$. Since the contours for $S^* = 1 - 4$ are similar, the results for $S^* = 2, 3$ are not displayed for brevity. There are two peaks corresponding to the rolling positions of the two shear layers from two sides of the downstream ($S^*$=1-6) and upstream ($S^*> 4$) cylinders (Fig. 12 (a-d)). As expected in extended-body flow ($S^*$= 1-4), the values of $\overline{u^*v^*}$ are almost zero between two cylinders. In addition, by changing the flow pattern to co-shedding $S^* \geq 5$, the shear stresses of the downstream cylinder are lower than those in extended-body flow ($S^*$= 1-4), (Fig. 13 (a-d, e)).

The streamwise distances between the peaks of $(\overline{u^*v^*})$ and the rear wall of cylinders are defined as $L_{uv}^*$ as seen in 13 (a, b). For the upstream cylinder, the $L_{uv}^*$ is only defined in the co-shedding flow pattern, and a higher $S^*$ leads to a higher $L_{uv}^*$ (Fig. 13 (e)). For the downstream cylinder in the extended-body flow ($S^* \leq 4$), due to the lack of vortex shedding between two cylinders the $L_{uv}^*$ increases with increasing $S^*$, but in the co-shedding flow pattern ($S^* \geq 5$) it diminishes (Fig. 13 (e)).

The total kinetic energy and the fluctuating stresses on the center line ($y^* = 0$) are shown in the extended-body flow, $S^* = 4$, (Fig. 13 (f)) and co-shedding pattern, $S^*$= 6, (Fig. 13 (g)) at $Re =$

150. In extended-body flow ($S^* \leq 4$), the fluctuating stresses are zero due to the absence of vortex shedding between two cylinders (Fig. 13 (f)). For $S^* \geq 5$ with vortex shedding from both cylinders, the fluctuating stresses and $KE^*$ between cylinders are non-zero and higher than those from the downstream cylinder (Fig. 13 (g)). All the fluctuating stresses for the downstream cylinder in $4 < S^*_{\text{cr}} \leq 5$ have an incremental jump due to changing the flow pattern from extended-body to co-shedding flow pattern.

### 4-3 Global parameters

#### 4-3-1 Local pressure and friction coefficients

Figure 14 (a, b) indicates the time-mean local pressure coefficient ($\overline{C}_p$) over both cylinders for various $S^*$ ($b_1^* = 1$, $b_2^* = 4$, $Re = 150$). The distributions of $\overline{C}_p$ on the front surface of the upstream cylinder (AB) are similar for all $S^*$ considered (Fig. 14 (a)). In $S^* \leq 4$ (extended-body flow), the absolute values of pressure coefficients in the upper/lower sides and back of the upstream cylinder decrease (or $\overline{C}_p$ becomes less negative) with increasing $S^*$ (Fig. 14 (a)). For $S^* \geq 5$ (the co-shedding flow), the $\overline{C}_p$ on the surfaces (DA-CD-BC) becomes more negative because of the vortex shedding (Fig. 14 (a)). This is also true for the downstream cylinder (Fig. 14 (b)).

A pair of symmetrical recirculation bubbles is formed between two cylinders for $S^* \leq 4$. The $\overline{C}_p$ on the front side of the downstream cylinder (AB) subsequently becomes negative (Fig. 14 (b)). But for the co-shedding flow with $S^* \geq 5$, the pressure on the (AB) is restored and becomes positive (Fig. 14 (b)) as the gap bubbles do not fill the space between the cylinders (Fig. 10 (b, c)). The $\overline{C}_p$ variations on the side faces of the downstream cylinder (BC, DA) depend on $S^*$. In the extended-body flow pattern ($S^* \leq 4$), the $\overline{C}_p$ values rise from (B) to the middle of the side and then become more negative until the corner (C). Whereas in co-shedding flow $S^* \geq 5$, the $\overline{C}_p$ values increase from (B) to the middle of the side and then it becomes relatively constant to (C) (Fig. 14 (b)).

In Fig. 14 (c, d), the distribution of the mean local friction coefficient $\overline{C}_f$ is shown for both cylinders ($S^* = 1\text{-}6$ and $Re = 150$). For the upstream cylinder, the $\overline{C}_f = 0$ at the stagnation point, then grows sharply as it approaches the front corner (B or A). After passing this corner, it decreases sharply because of the flow separation towards the rear corner (C or D) (Fig. 14 (c)). The $\overline{C}_f$ dependence on $S^*$ for the downstream cylinder is different. For $S^* \leq 4$ (*EBF*), the $\overline{C}_f$ on the front face becomes negative with a zero value at the stagnation point (Fig. 14 (d)). Due to the reattachment point on the side faces ($\overline{C}_f = 0$) at a position of about $0.27b_2^*$ from the front corners, the $\overline{C}_f$ sign changes from negative to positive and then steadily increases (Fig. 14 (d)). But for $S^* \geq 5$ (*CSF*), the $\overline{C}_f$ becomes positive on the front face (Fig. 14 (d)). Then, it decreases to a relatively constant value in the order of zero along the side faces because of the vortex shedding formation.

### 4-3-2 Total pressure and frictional drag coefficients

Figure 15 represents the profiles and contours of the time average of total drag ($\overline{C}_d$), pressure drag ($\overline{C}_{dp}$), and frictional drag ($\overline{C}_{df}$) coefficients for various $Re$ and $S^*$ for both cylinders ($b_1^* = 1$, $b_2^* = 4$). The line with the circle in Fig. 15 (c, d) shows the $Re_{cr}$, and the dashed lines divide the time-mean flow patterns. Fig. 15 (a) shows that the $\overline{C}_d$ of both cylinders at $S^* \leq 4$ steadily decreases with increasing $Re$. For both cylinders at $S^* = 5$, a jump is observed in $\overline{C}_d$ (Fig. 14 (a, b)) at $Re = 70\text{-}75$ where the vortex shedding occurs, and flow pattern changes from steady extended-body to co-shedding (see also Fig. 8 (a)). Fig. 15 (a-d) demonstrates that the $\overline{C}_d$ values of both cylinders in the co-shedding flow pattern ($S^* \geq 5$) are larger than those in the extended-body flow pattern ($S^* \leq 4$). In the *EBF* pattern for the downstream cylinder ($b_2^* = 4$), the symmetrical vortices in the gap cause a strong suction on the front face of the downstream cylinder. The negative pressure on this face makes the drag force of this cylinder negative for $S^* \leq 4$ but it depends on $Re$, e.g. at $S^* = 1$ for $Re > 100$ (Fig. 15 (b, d)).

Since $\overline{C}_d = \overline{C}_{dp} + \overline{C}_{df}$, scrutiny of $\overline{C}_{dp}$ and $\overline{C}_{df}$ comes with further drag variation clarification. The $\overline{C}_{dp}$ behavior is almost similar to $\overline{C}_d$ for both cylinders (Fig. 15 (e, f)) because the $\overline{C}_{df}$ is generally smaller than $\overline{C}_{dp}$, although $\overline{C}_{df}$ at low Reynolds numbers (dependent on $Re$ and $S^*$) has a noticeable contribution [20]. The $\overline{C}_{df}$ decreases with $Re$ for both cylinders (Fig. 15 (g, h)). By increasing $Re$, the separation points of the upstream cylinder shift from rear corners to front ones, and the recirculation bubbles are formed on its side faces. This makes $\overline{C}_{df}$ negative for $S^* \leq 4$ at $Re = 150$ (e.g., see Fig. 14 (c), Fig. 15 (g)). For the downstream cylinder, there is neither the separation bubble nor the flow reattachment. Therefore, the $\overline{C}_{df}$ remains positive (e.g., Fig. 15 (h)).

### 4-3-3 RMS lift and drag coefficients

How the $C_{lrms}$ and $C_{drms}$ are dependent on $Re$ and $S^*$ is illustrated in Fig. 16 ($b_1^* = 1$, $b_2^* = 4$). These two coefficients show the force fluctuation level caused by vortex shedding and mainly increase with increasing $Re$. They are zero in the steady flow, e.g. $Re < 100$ for $S^* \leq 4$ ($b_1^* = 1$, $b_2^* = 4$), see Fig. 8 and 16. In the extended-body flow pattern for $S^* \leq 4$, the $C_{lrms}$ for the upstream cylinder is in order of zero, while for the downstream cylinder is not zero but it is not much (Fig. 16 (a, b)). At $Re > 120$, it increases slightly. However, the $C_{lrms}$ increases with increasing $Re$ for both cylinders in a co-shedding flow pattern, e.g. at $S^* = 6$. This increase is more significant for the downstream cylinder (Fig. 16 (a, b)).

The $C_{drms}$ is zero in extended-body flow ($S^* \leq 4$) for the upstream cylinder. Fig.16 (c) shows the $C_{drms}$ raises as the strength of vortices increases at higher $Re$ for both cylinders. The impinging gap vortices produce a lower fluctuation force on the upstream cylinder compared to the downstream cylinder. Hence, $C_{drms}$ is higher on the downstream cylinder (Fig. 16 (c)). For $S^* \geq 5$ with vortex shedding and increasing $S^*$ ($Re \leq 100$), the $C_{drms}$ grows for the downstream cylinder, while its variation is insignificant for $Re > 100$.

As mentioned, the $L_f^*$ has an inverse relationship with the RMS of drag and lift for a single cylinder [46, 47]. While, for two tandem cylinders, it depends on the $S^*$ and flow pattern. A higher $L_f^*$ and $S^*$ corresponded to lower $C_{lrms}$ for the upstream cylinder in the co-shedding flow. But, the $L_f^*$ has a direct relationship with $C_{lrms}$ and $S^*$ for the downstream cylinder in extended-body flow (Fig. 12 (a-d, h) and Fig. 16 (a, b)).

**4-3-4 Vortex shedding frequency**

Figure 17 illustrates the Strouhal number (*St*) as a function of *Re* and $S^*$ ($b_1^* = 1$, $b_2^* = 4$). The contour incorporates the $Re_{cr}$ and the time-mean flow pattern boundaries. The *St* is determined from the fast Fourier transform (FFT) of the fluctuating lift. As expected, the *St* = 0 for cases with $Re < Re_{cr}$ and the upstream cylinder in the extended-body flow pattern ($S^* \leq 4$). For $S^* \leq 4$, the *St* associated with the downstream cylinder increases with increasing *Re* and decreases insignificantly with $S^*$. Nonetheless, the *St* is the same for both cylinders and increases with increasing *Re* and $S^*$ for $S^* \geq 5$ (i.e., co-shedding flow pattern), which can be ascribed to elongated $L_e^*$ with $S^*$. For the single cylinders, the $L_e^*$ has an inverse relationship with *St* [14, 43]. A comparison of Figs. 12 (h) and 17 (a) reveal that the $L_e^*$ and *St* variations with $S^*$ similarly have opposite trends when $S^* \leq 4$ (i.e. extended-body flow pattern). But, for the co-shedding flow pattern, the $L_e^*$ and *St* both increase with $S^*$. Equations (8) and (9) reflect our simulation results for the vortex shedding frequency, showing how *St* varies with *Re* and $S^*$ ($b_1^* = 1$, $b_2^* = 4$). These equations respectively deviate ~4.5% and 3% from data in Fig. 17 at the most.

$$St = (0.05 + \frac{0.001}{S^* + 0.07}) \ln(Re) - 0.1263, \qquad S^* \leq 4, \qquad \text{For Cyl 2,} \qquad (8)$$

$$St = \left(\frac{0.5}{S^*} + 1.9\right) \times 10^{-6} Re^2 + \left(\frac{0.00075 S^*}{S^* + 0.05}\right) Re + 0.081, \qquad S^* \geq 5, \qquad \text{For Cyl 1, 2,} \qquad (9)$$

**4-4 Hysteresis phenomena**

The dependence of flow characteristics on the approaching manner of *Re* and/or $S^*$ into the target flow condition (i.e. the flow history) is called hysteresis in the context of flow over cylinders in arrangements. In other words, hysteresis is a phenomenon of the flow resisting changing the flow pattern. It is mainly observed in a range around the *Re* or $S^*$ thresholds, in which the flow pattern transits. It is numerically investigated by applying different initial conditions, where two different aerodynamic responses are observed in a given $S^*$ and *Re* [19, 20, 23, 25]. The experimental studies gradually increase and decrease the $S^*$, for a given *Re* [22, 24]. The $S^*$ and/or *Re* range, in which this phenomenon is observed, is known as the hysteresis limit.

The hysteresis here is studied in various $S^* = 1-6$ at $Re = 150$ ($b_1^* = 1$, $b_2^* = 4$). Three different initial conditions are applied for a given $S^*$ to find the hysteresis limit and are named A, B, and C conditions. In case A, the simulation is started with a fluid at rest. In case B, an already simulated case for a larger $S^*$ (e.g. the flow field for a case with $S^* = 5$ is mapped into and initialized the case with $S^* = 4$) serves as the initial condition, while *Re* is fixed for both of them. The procedure is reversed in case C, compared with case B, i.e., the data of a smaller $S^*$ initializes the case with a larger $S^*$. The hysteresis in this study is observed at $3.5 \leq S^* \leq 4.5$. Two distinct responses in the hysteresis limit are seen (Fig. 18) and referred to here as mode 1 (extended-body flow pattern or *EBF*) and mode 2 (co-shedding flow pattern or *CSF*), i.e., cases A and C come with the same results.

Figure 19 shows the variations of total drag coefficient ($\overline{C}_d$), RMS lift coefficient ($C_{lrms}$), and Strouhal number (*St*), in which the hysteresis limit (*HL*) has been marked. In *HL*, the two branches correspond to Mode 1 and Mode 2. In other words, the flow pattern is either the extended-body flow (Mode 1 or *EBF*) or co-shedding flow (Mode 2 or *CSF*) outside the *HL* region, regardless of the initial condition. However, the initial condition ascertains which flow pattern to emerge in the *HL*.

The $\overline{C}_d$ gradually decreases for the upstream cylinder and moderately increases for the downstream cylinder with increasing $S^*$ for Mode 1. An abrupt increase is then observed in the $\overline{C}_d$ of both cylinders in the *HL* due to shifting flow mode (Fig. 19 (a, b)). The drag coefficients associated with Mode 1 are smaller than those of Mode 2 in the *HL* (Fig. 19 (a, b)). The $\overline{C}_d$ of both cylinders gradually increases with $S^*$ in Mode 2 (*CSF*). It is worth noting that the $\overline{C}_d$ values for a given $S^*$ are larger for the upstream cylinder in both modes (Fig.18 (a-b)). RMS lift coefficients ($C_{lrms}$) are illustrated in Fig. 19 (c, d) for both modes and cylinders. The $C_{lrms}$ values for both cylinders are remarkably larger in Mode 2 (Fig. 19 (c, d)), where vortex shedding occurs from both cylinders (*CSF*). Similarly, the *St* is noticeable where the vortex shedding occurs (Fig. 19 (e)). However, the trend is a bit different from $C_{lrms}$.

Although the hysteresis for the tandem cylinders with square and rectangular cross-sections is reported for the first time in the literature, a comparison of results with similar studies over two tandem square cylinders is worthy. The *HL* range was reported as $2.5 \leq S^* \leq 4.75$ for two tandem square cylinders at *Re* = 150 [20, 23, 26]. Consequently, an increase in the downstream cylinder aspect ratio results in an *HL* decrement.

**4-5 Discussion**

The influence of *Re*, $b_2^*$, and $S^*$ on the flow features was investigated in the previous sections. We here aim to take a closer look into the results to relate the flow behavior and its correlation with the engineering parameters. Figure 20 represents the correlation of the time-mean flow pattern (*MFP*; *I*, *II*, *III*) and instantaneous flow pattern (*IFP*; i.e., extend-body flow or *EBF* and co-shedding flow or *CSF*) discussed in previous sections. Table 7 (a, b), on the other hand, renders qualitative variations of global characteristics, fluctuating stress, and kinetic energy in different flow patterns ($b_1^* = 1$, $b_2^* = 4$, $S^* = 1 - 6$, *Re* = 30-150).

To compare the flow physics and correlate the aerodynamic characteristics and flow patterns, two Reynolds numbers of 70 and 150 associated with the steady and unsteady flows (i.e. before and after the onset of vortex shedding), are considered. At *Re* = 70 and $1 \leq S^* \leq 4$, the time-mean flow is *I* (steady) and the flow separation takes place from the trailing edge of the upstream cylinder, similar to that occurs from a square cylinder up to *Re* = 100 [9]. The flow is then attached to the side surfaces of the downstream cylinder and separated again from the downstream cylinder trailing corners. The time-mean flow *I* at *Re* = 70 corresponds to the steady flow for $S^* \leq 5$ and unsteady flow for $S^* = 6$ (co-shedding flow pattern), see Figs. 8, 9 and 20 (a, b). Thus, an increase of $S^*$ changes the steady flow to co-shedding flow although the separation still takes place from the rear edge of the cylinders (Fig. 20 (b)).

At $S^* = 1$-3 and *Re* = 150, the time-average streamlines are separated from the front edges of the upstream cylinder and stick to the side surfaces of the downstream cylinder (Fig. 20 (c)). The *EBF* and *III* flow patterns are found from instantaneous and mean flows, respectively, see Figs. 8 and 9. Due to the hysteresis phenomena both *EBF* and *CSF* were observed at $S^* = 4$, *Re* = 150. These two instantaneous flow patterns are observed in Fig. 20 (d) and are denoted as Mode 1 and Mode 2, respectively. The corresponding mean-flow patterns are *II* and *III*, respectively (Figs. 9, 10, and 20 (c)). At $S^* = 6$ and *Re* = 150, the instantaneous and mean flow patterns are *CSF* and *II*, respectively, see Fig. 20 (e).

To obtain a qualitative relationship between the aerodynamic characteristics and flow patterns; two cases are considered. First at a constant *Re* = 150, $S^*$ varies from 1-6 (Table 7 (a)). Second, *Re* changes from 30 to 150 at a constant $S^* = 6$ (Table 7 (b)).

Table 7 (a) indicates that the vortex shedding frequency, force fluctuations, shear stress, and kinetic energy for the upstream cylinder are null, while the length and width of the vortices behind the upstream cylinder ($L_r^*$, $W_r^*$) increase and consequently, the $\overline{C}_{dp}$ and $\overline{C}_d$ decrease (also see Figs. 11 and 15) when $S^*$ changes from 1 to 4 at *Re* = 150 (*EBF*, mean flow patterns *II* and *III*). For the downstream cylinder by changing $S^*$ from 1 to 4, the drag coefficient and force fluctuations increase, while the other parameters in Table 7 (a) decrease (*CSF*, mean flow patterns *II* and *III*). For $S^* = 5, 6$, the co-shedding flow pattern occurs and variations of parameters are not similar to those for $S^* = 1$-4 (Table 7 (a)). For example, the $KE^*$, $L_r^*$, and *St* increase for the upstream cylinder, while the $\overline{u^*v^*}$ and RMS lift and drag decrease. For the downstream cylinder, the $L_r^*$ decreases, while $\overline{C}_d$, $C_{drms}$, *St*, $W_r^*$, and $KE^*$ increase.

In the second case, because the cylinder gap-spacing is considered constant ($S^* = 6$) and the Reynolds number increases (*Re* = 30-150), both steady (*Re* < *Re*$_{cr}$) and unsteady (*Re* > *Re*$_{cr}$) flow behavior appears (Table 7 (b)). Thus, at $S^* = 6$, with increasing *Re* = 30-150 and mean flow patterns *I, II,* unsteady *MFP*, and *CSF*, an increment was observed in the RMS lift and drag, vortex shedding frequency, fluctuating stress, and kinetic energy for both cylinders. The $L_r^*$ variation has different behavior and decreases in *EBF* for the upstream cylinder and increases for the downstream cylinder, and it is related to the base pressure of the cylinders. Only $W_r^*$ remains unchanged for the upstream cylinder in the *CSF* (Table. 7 (b)). The $\overline{C}_d$ reduces for both cylinders in any case (Fig. 15 and Table 7 (b)).

Table 7. Qualitative variations of engineering parameters, fluctuating stress, and kinetic energy in different flow patterns for various $S^*$ and $Re$ ($b_1^* = 1$, $b_2^* = 4$), (a): $S^* = 1 - 6$, $Re$ =150, (b): $S^* = 6$, $Re$ =30-150.

| IFP | Cyl | MFP | $\bar{C}_d$ | $C_{lrms}$ | $C_{drms}$ | St | $L_r^*$ | $W_r^*$ | $\overline{u^*v^*}$ | $KE^*$ |
|---|---|---|---|---|---|---|---|---|---|---|
| (a) | | | | $S^*$= 1-6 | | | | $Re$=150 | | |
| EBF, $S^*$=1-4 | 1 | II, III | Decrease | 0 | 0 | 0 | Increase | Increase | 0 | 0 |
| | 2 | II, III | Increase | Decrease | Decrease | Decrease | Decrease | Decrease | Decrease | Decrease |
| CSF, $S^*$= 5, 6 | 1 | II, III | Unchanged | Decrease | Increase | Increase | Unchanged | Decrease | Decrease | Increase |
| | 2 | II, III | Increase | Unchanged | Increase | Decrease | Increase | Unchanged | Unchanged | Increase |
| (b) | | | | $S^* = 6$ | | | | $Re$ = 30-150 | | |
| EBF, $Re < Re_{cr}$ | 1 | I, II | Decrease | Increase | Increase | Increase | Decrease | Increase | Increase | Increase |
| | 2 | I, II | Increase | Increase | Increase | Increase | Increase | Increase | Increase | Increase |
| CSF, $Re > Re_{cr}$ | 1 | I, II | Increase | Increase | Increase | Increase | Decrease | Unchanged | Increase | Increase |
| | 2 | I, II | Decrease | Increase | Increase | Increase | Increase | Increase | Increase | Increase |

## 5- Conclusions

The results of flow around tandem Sharp-edged cylinders of diverse cross sections at different cylinder gap-spacing $S^*$ =1-6 and $Re$ = 30-150 were investigated. The aspect ratio of the downstream rectangle cylinder ($b_2^* = 1 - 4$) was also studied. The ranges of critical Reynolds number ($Re_{cr}$) that vortex shedding occurs and the flow changes from steady to unsteady were determined for both cylinders for various $b_2^*$ and $S^*$. Besides, the instantaneous (extended-body flow (*EBF*) and co-shedding flow (*CSF*)) and time-mean flow (*I, II, III*) structure along with the details of the flow (aerodynamic parameters and fluctuating stresses) around the cylinders were analyzed.

The $Re_{cr}$ changes for two tandem square $b_1^*$= 1 and rectangular $b_2^*$= 4 cylinders as a second and third-order polynomial (Eqs. 6 and 7), the $Re_{cr}$ decreases from 128±2 to 50±2 with $S^*$ increasing from 1 to 6. A smaller $Re$ is thus required for a larger $S^*$ to have the onset of vortex shedding ($b_1^* = 1, b_2^* = 4$). However, at $S^* = 4$ for two tandem square cylinders ($b_1^* = b_2^* = 1$), $Re_{cr} = 78±2$ is the same for both cylinders, but $Re_{cr} = 165±2, 145±2$ and $93±2$ for $b_2^* = 2, 3$ and $4$, respectively. So, at a fixed $S^* = 4$, with the aspect ratio's growth of the downstream cylinder $b_2^*$ =1-4, the $Re_{cr}$ for the downstream cylinder first increases up to $b_2^* = 2$ and then decreases. The elongated streamwise length of the downstream cylinder stabilizes the flow like a splitter plate. In $S^* = 4$ by increasing the downstream cylinder aspect ratios, the vortex shedding is present for both cylinders at $b_2^* = 1$ (*CSF*), suppressed for both cylinders up to $Re_{cr} = 165\pm2$ ($b_2^* = 2$) and then stopped for the upstream cylinder at $b_2^* = 3, 4$ for $Re_{cr} < Re \leq 150$ (*EBF*).

Two instantaneous flow patterns are apperceived and in the ranges of *Re*, *S\** and *b\** examined. In extended-body flow (*EBF*), the shear layers separated from the upstream cylinder stick to the surface of the downstream cylinder, and vortex shedding just occurs from the downstream cylinder. This flow pattern has been observed at $S^* \leq 4$, and $Re = Re_{cr} - 150$ for two tandem cylinders ($b_1^* = 1, b_2^* > 2$). In the co-shedding flow (*CSF*), the vortex shedding takes place simultaneously between two cylinders and behind the downstream cylinder. This flow pattern can be seen in $Re = Re_{cr} - 150$ at $S^* \geq 5$ ($b_1^* = 1, b_2^* = 1, 4$). For $b_1^* = 1, b_2^* = 2$, flow over both cylinders becomes steady up to $Re = Re_{cr} = 165\pm2$.

Three distinct flow patterns (*I*, *II*, and *III*) of mean flow are identified. In pattern *I* (rear-corner separated flow) the flow is completely parallel to the upper and lower surfaces of the upstream cylinder and no reattachment or separation bubble is created on the cylinder surfaces. In pattern *II* (separation bubble flow), flow separates from the front side of the upstream cylinder and a pair of separation bubbles is formed on the upper and lower surfaces of the upstream cylinder and then, the reattachment point is created on those surfaces. In pattern *III* (front-corner separated flow) the separation bubble on the upper and lower surfaces covers almost the entire surface of the upstream cylinder and no reattachment point is observed. A saddle point emerges above the upstream cylinder rear corners. For $b_1^* = 1, b_2^* = 4$, mean flow pattern *I*, prevails at $S^* \leq 4$ in $Re$ = 30-95, pattern *II* appears at $S^* \geq 5$ in $Re$ = 120-150 and pattern *III* emerges at $S^* \leq 2$ in $Re$ = 120-150. Pattern *III* has not been observed for $S^* = 5, 6$, while it occurs for $S^* \leq 4$ and $Re > 105$.

Fluctuation stresses perpendicular to the flow (normal stress) are stronger than those in the flow direction and are more noticeable in unsteady flow between two cylinders. In both mean flow patterns (*II* and *III*), the ($\overline{v^*v^*}$) is the maximum oscillating stress, which contributes more to the total kinetic energy. All the fluctuating stresses for the downstream cylinder in $4 < S_{cr}^* \leq 5$ and the range of changing flow from quasi-steady to unsteady have an incremental jump.

Aerodynamic parameters are completely influenced by *Re*, *S\**, downstream cylinder aspect ratios ($b_2^*$), and flow patterns when the *Re* increases and the instantaneous flow pattern changes from extended-body (*EBF*) to co-shedding (*CSF*), an increase in all aerodynamic parameters was observed for the two cylinders. Aerodynamic parameters, except base pressure behind the cylinders, after variation in the flow pattern and incremental jump, continue their previous behavior with the increase of Reynolds number. Recirculation bubbles size behind the cylinders as well as the length of the vortex formation have an inverse relationship with base pressure behind both cylinders.

The results of this research authenticate the hysteresis limit for flow around two tandem cylinders ($b_1^* = 1, b_2^* = 4$) occurs in cylinder gap-spacing $3.5 \leq S^* \leq 4.5$ ($Re = 150$). In the hysteresis limit, two branches are observed. Each branch corresponds to an instantaneous flow pattern. The lower branch indicates extended-body flow where there is no vortex shedding between two cylinders (Mode 1). While the upper branch represents a co-shedding flow pattern (Mode 2). Hence, the hysteresis limit depends on the Reynolds number, cylinder gap spacing, and aspect ratio of the upstream cylinder. It is noteworthy that in all situations and for both cylinders, the drag coefficient, $C_{lrms}$, and Strouhal number in Mode 2 are higher than those of Mode 1.

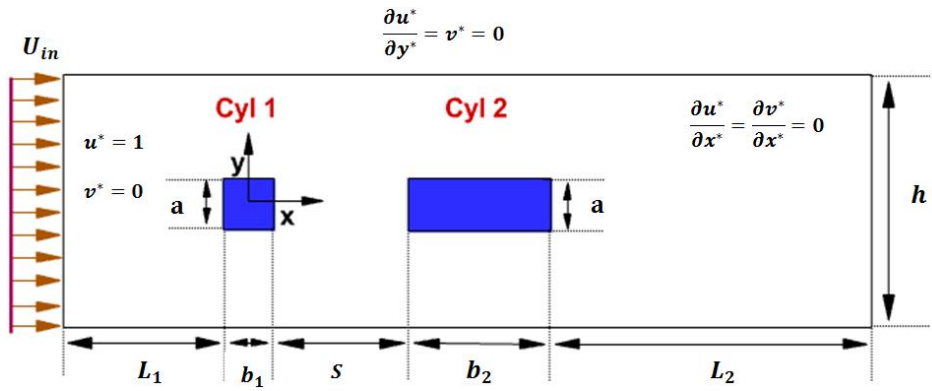

Figure 1. Computational domain and boundary conditions for the problem under consideration (Cyl1: upstream cylinder: Cyl2: downstream cylinder).

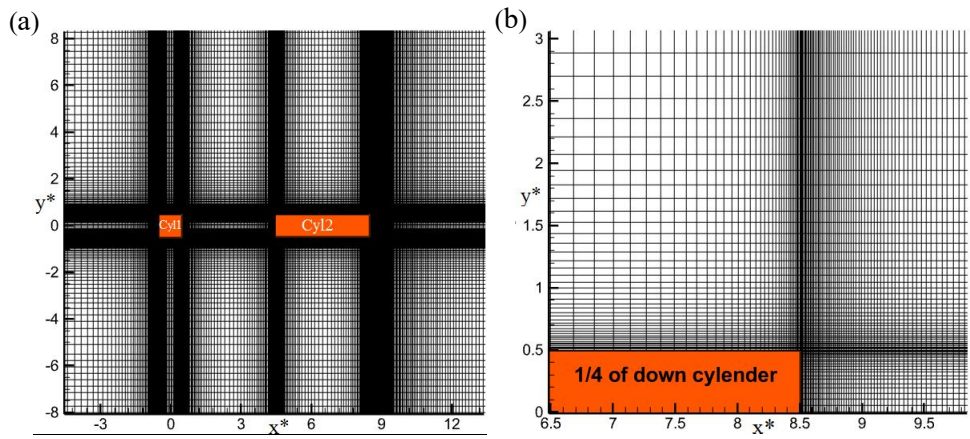

Figure 2. The close-up grid ($b_1^* = 1, b_2^* = 4, S^* = 4$) in (a) around two cylinders and (b) around 1/4 downstream cylinder.

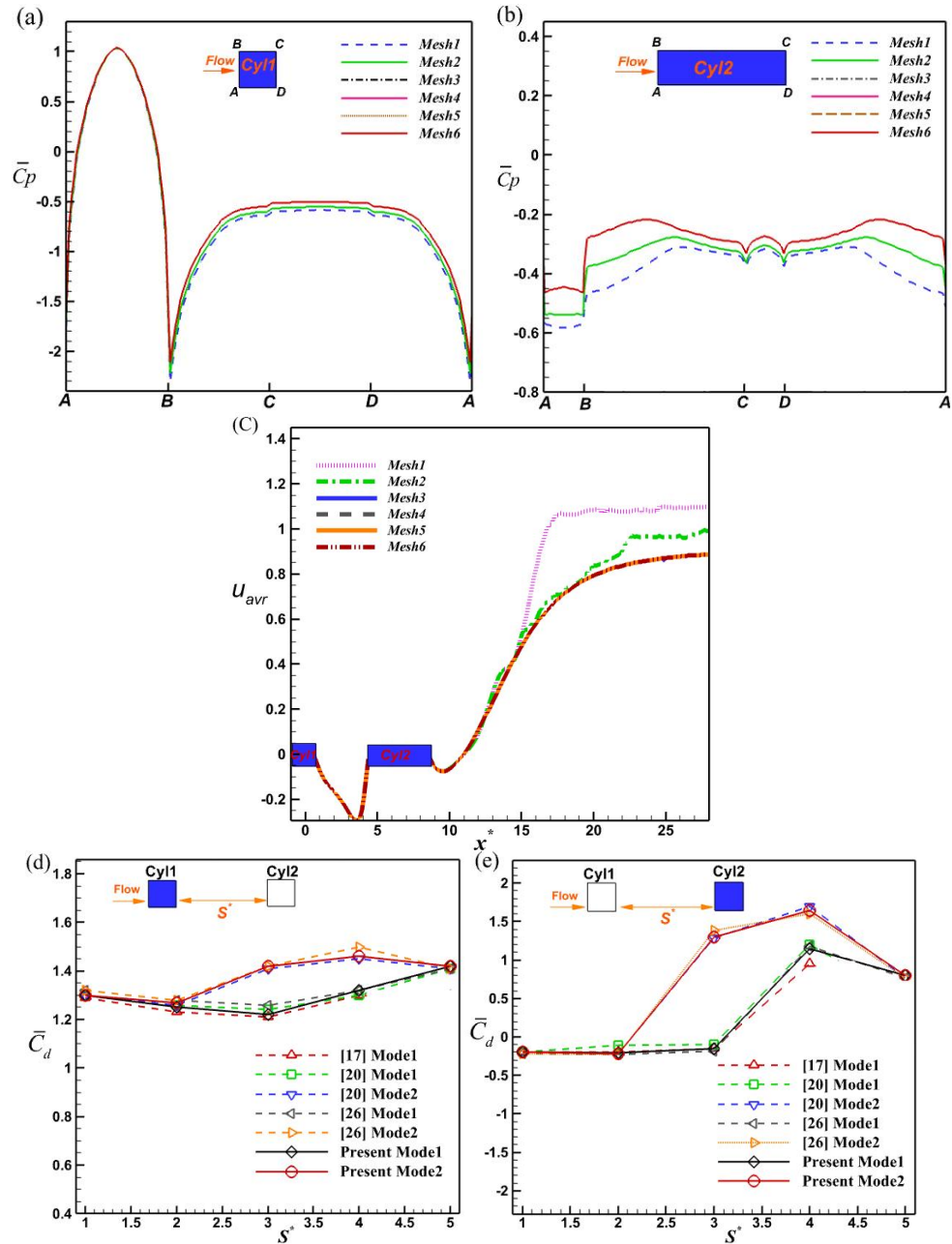

Figure 3. Grid study and validation test ($Re$ = 150). The local pressure coefficient on the cylinders surfaces (a) Cyl1 ($b_1^* = 1$), (b) Cyl2 ($b_2^* = 4$), (c) the mean velocity on the centerline (y=0) for various grids ($S^*$ = 4). $\overline{C}_d$ comparison with [17, 20, 26] as a validation test for two tandem square cylinders at different $S^*$ (d) Cyl1 ($b_1^* = 1$), (e) Cyl2 ($b_2^* = 1$).

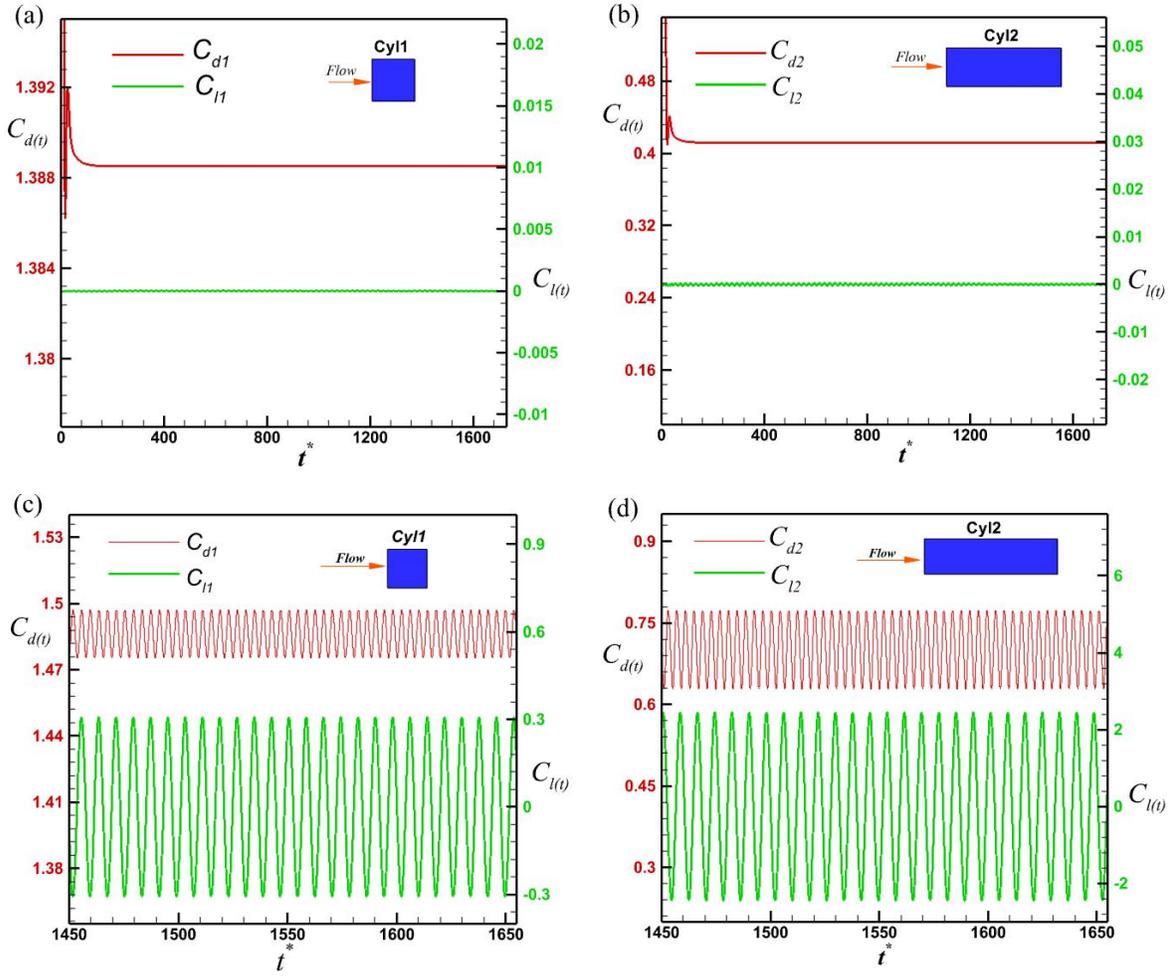

Figure 4. The lift and drag coefficient signals of both cylinders for (a, b) steady ($S^* = 5$ and $Re = 70$) and (c, d) unsteady flows ($S^* = 4$ and $Re = 75$).

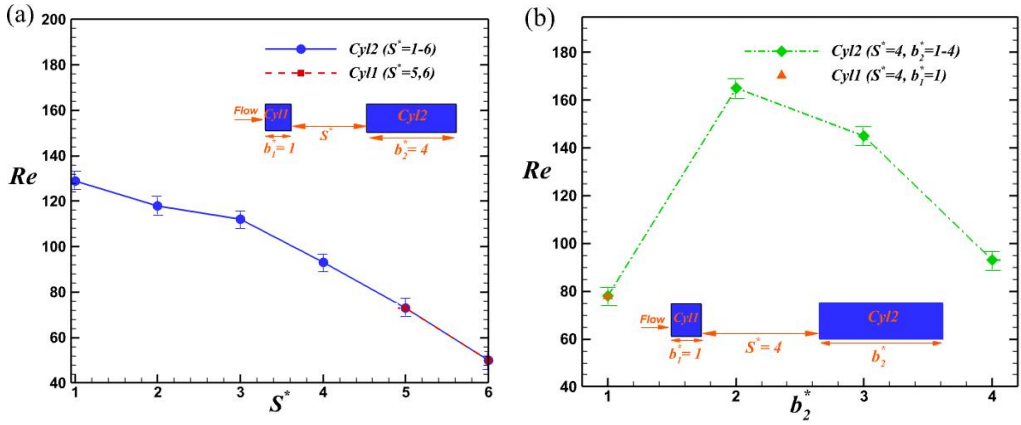

Figure 5. The variation of critical Reynolds numbers ($Re_{cr}$) for various (a) $S^* = 1 - 6$ ($b_1^* = 1$, $b_2^* = 4$) and (b) $b_2^* = 1 - 4$ ($S^* = 4$, $b_1^* = 1$).

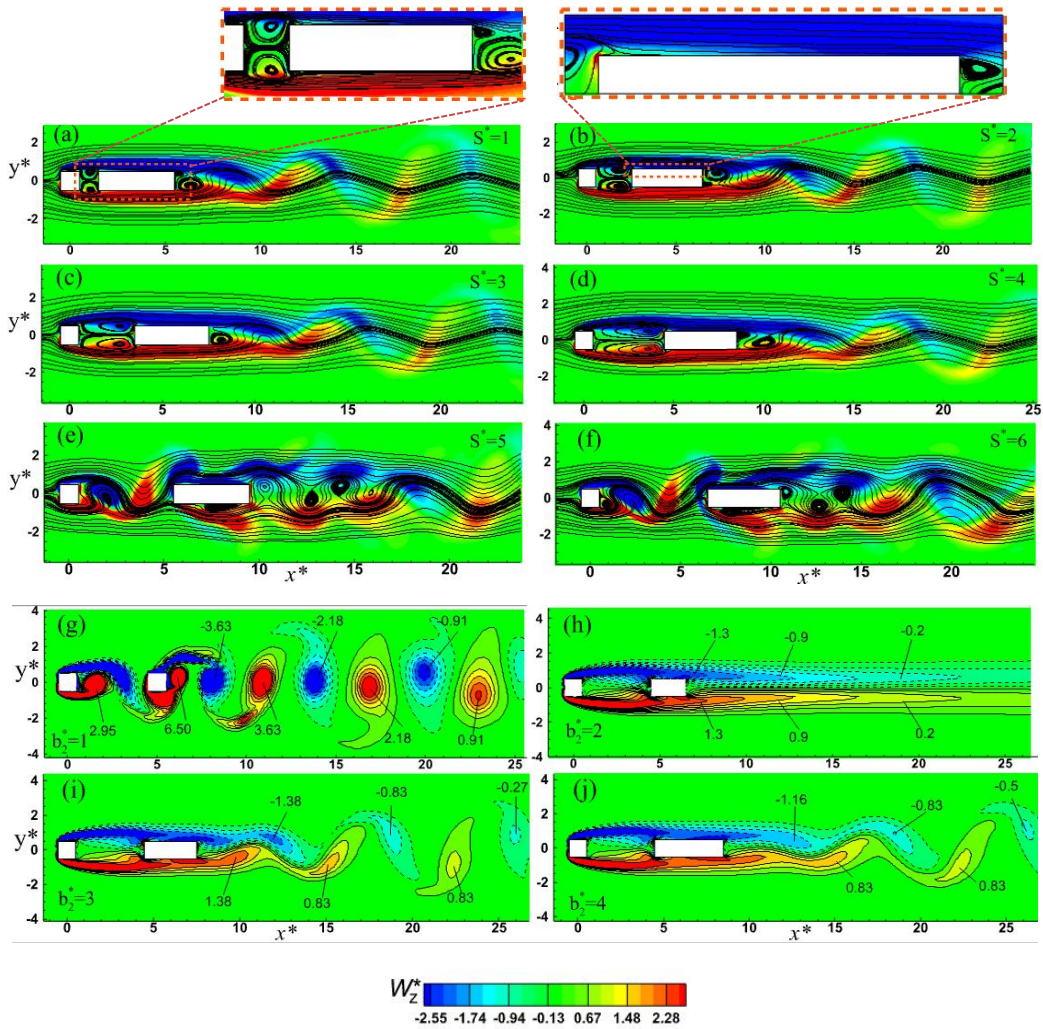

Figure 6. (a-f) Instantaneous streamlines colored with vorticity for various $S^*$ at $b_2^* = 4$ (g-j) instantaneous vorticity for different $b_2^*$=1-4 at $S^* = 4$ ($Re$ = 150 and $b_1^* = 1$).

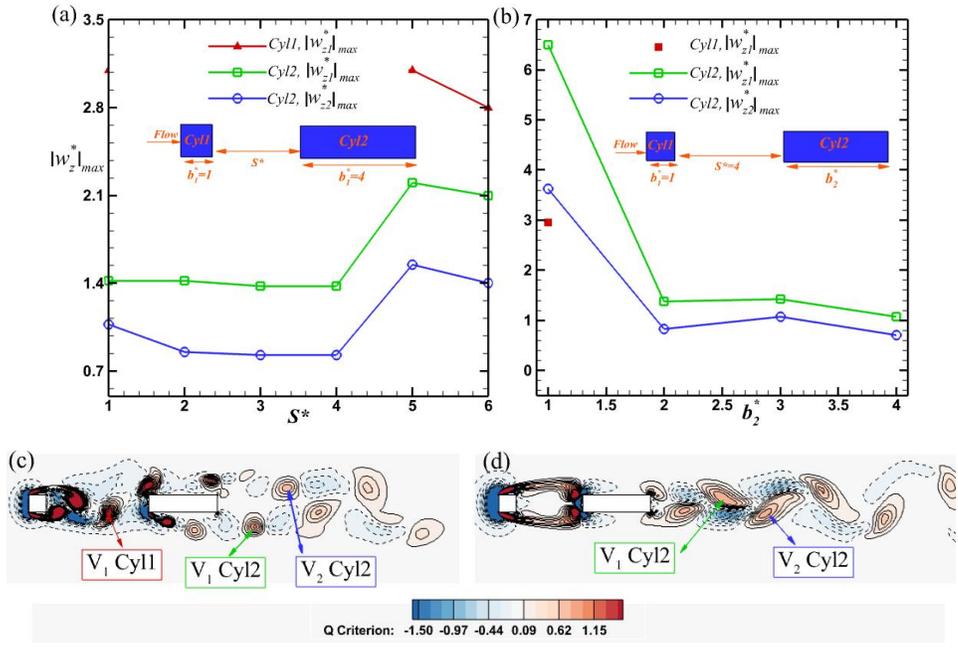

Figure 7. The maximum vorticity magnitude for various (a) $S^*$ ($Re = 150$, $b_2^* = 4$) and (b) $b_2^*$ ($Re=170$, $S^*=4$), $|\omega_z^*|_{max}$. The detached vortices identified with the Q criterion behind the upstream and downstream cylinders discussed in (a, b) have been marked in subplots c (*CSF*) and d (*EBF*).

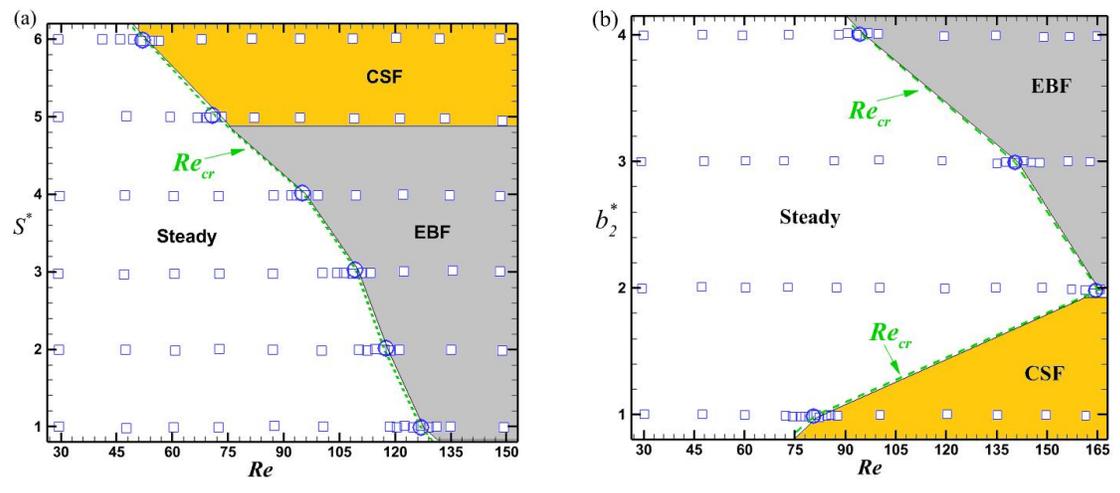

Figure 8. View of all simulations performed (a) for various $S^*$ and $Re$ ($b_1^* = 1$, $b_2^* = 4$), (b) for various $b_2^*$ and $Re$ ($b_1^* = 1$, $S^* = 4$), wherein these flow maps *EBF* and *CSF* stand for extended-body flow and co-shedding flow, respectively.

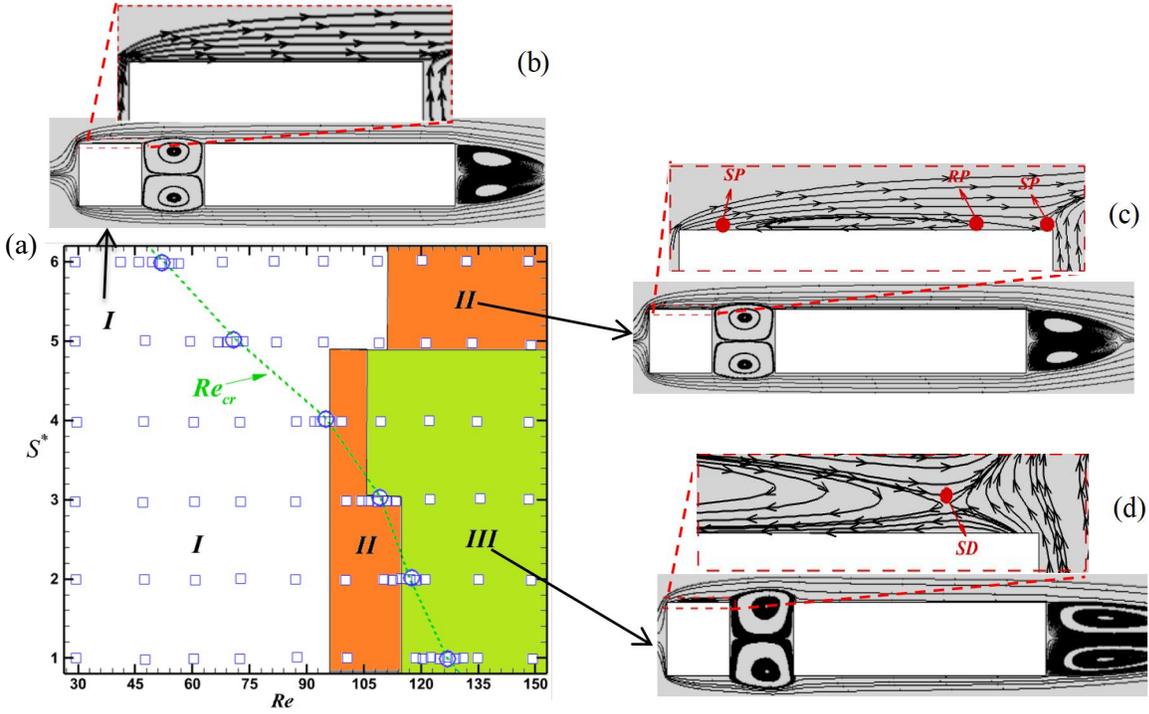

Figure 9. (a) Marginal time-mean flow map at $S^*$-$Re$ plane. The simulated cases with $b_1^* = 1$, $b_2^* = 4$, at a wide range of $S^*$ and $Re$ in the present study have been marked with a blue hollow square. The green dashed line with a hollow circle indicates $Re_{cr}$. The mean streamlines of the three time-mean flows are shown for patterns (b) *I* (white background in (a)), (c) *II* (orange background in (a)), (d) *III* (green background in (a)). *SP*: Separation point, *RP*: Reattachment point, *SD*: Saddle point.

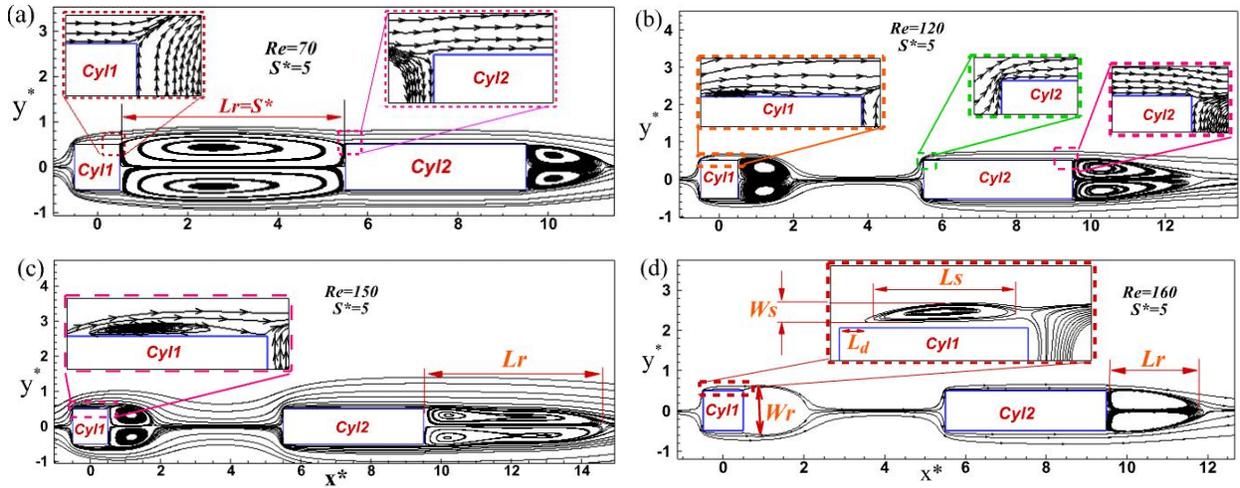

Figure 10. Mean flow (a) pattern *I*, (b, c) pattern *II* and (d) pattern *III* ($b_1^* = 1$, $b_2^* = 4$, $S^* = 5$ and $Re = $ 70-150). The parameters of $L_r$, $W_r$, $L_s$, $W_s$, and $L_d$ are shown in (a, d).

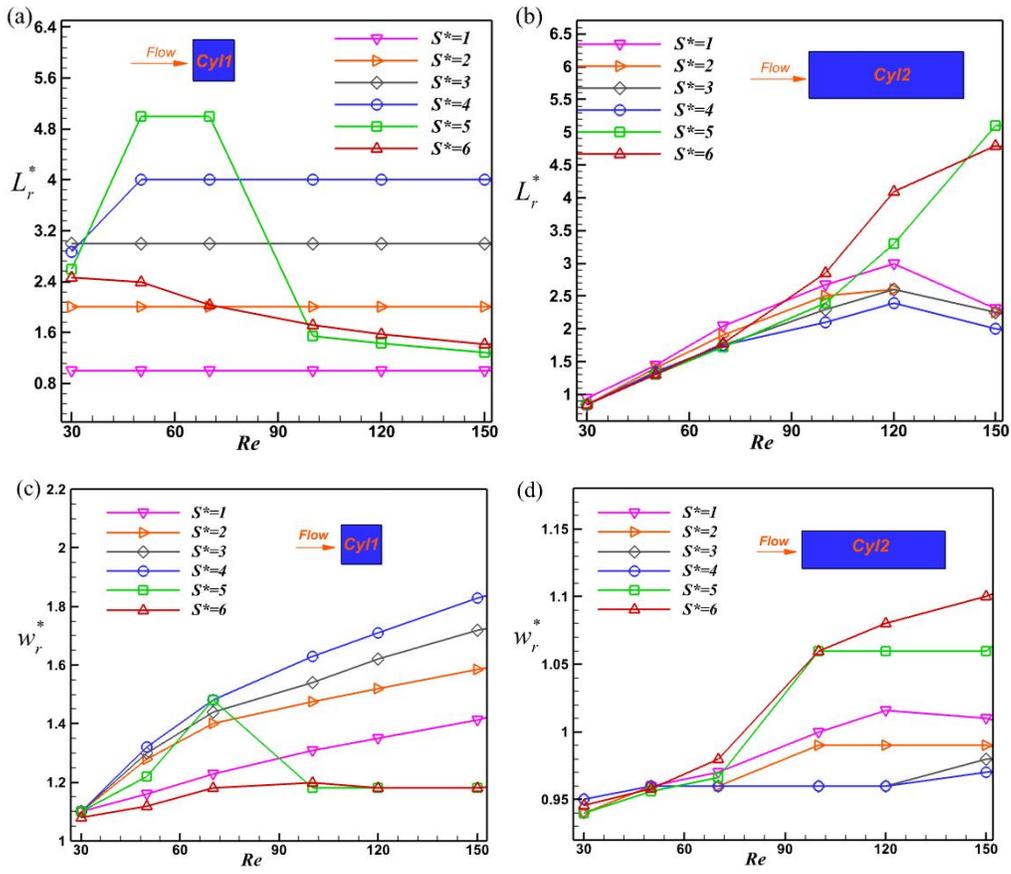

Figure 11. (a, b) the bubble length $L_r^*$, and (c, d) bubble width $W_r^*$ for both cylinders at various $Re$ and $S^*$ ($b_2^* = 4$), see also Fig. 10(a, d).

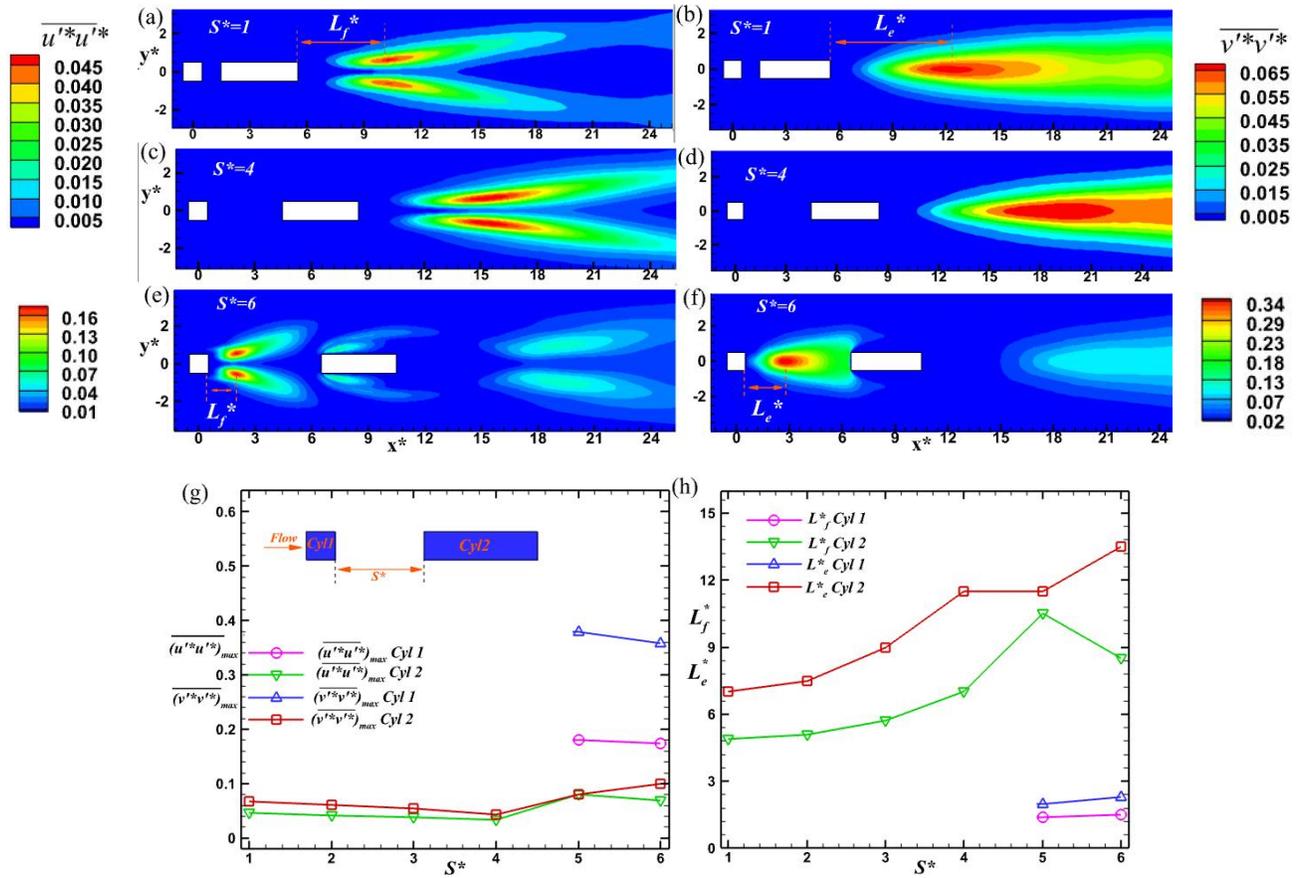

Figure 12. (a, c, e) The $(\overline{u'^{*}u'^{*}})$ and (b, d, f) $(\overline{v'^{*}v'^{*}})$ contours for various $S^*$. (g) The variation of maximum values, i.e., $(\overline{u'^{*}u'^{*}})_{max}$ and $(\overline{v'^{*}v'^{*}})_{max}$ (h) $L^*_e$ and $L^*_f$ with $S^*$ ($b_1^* = 1$, $b_2^* = 4$, and $Re = 150$).

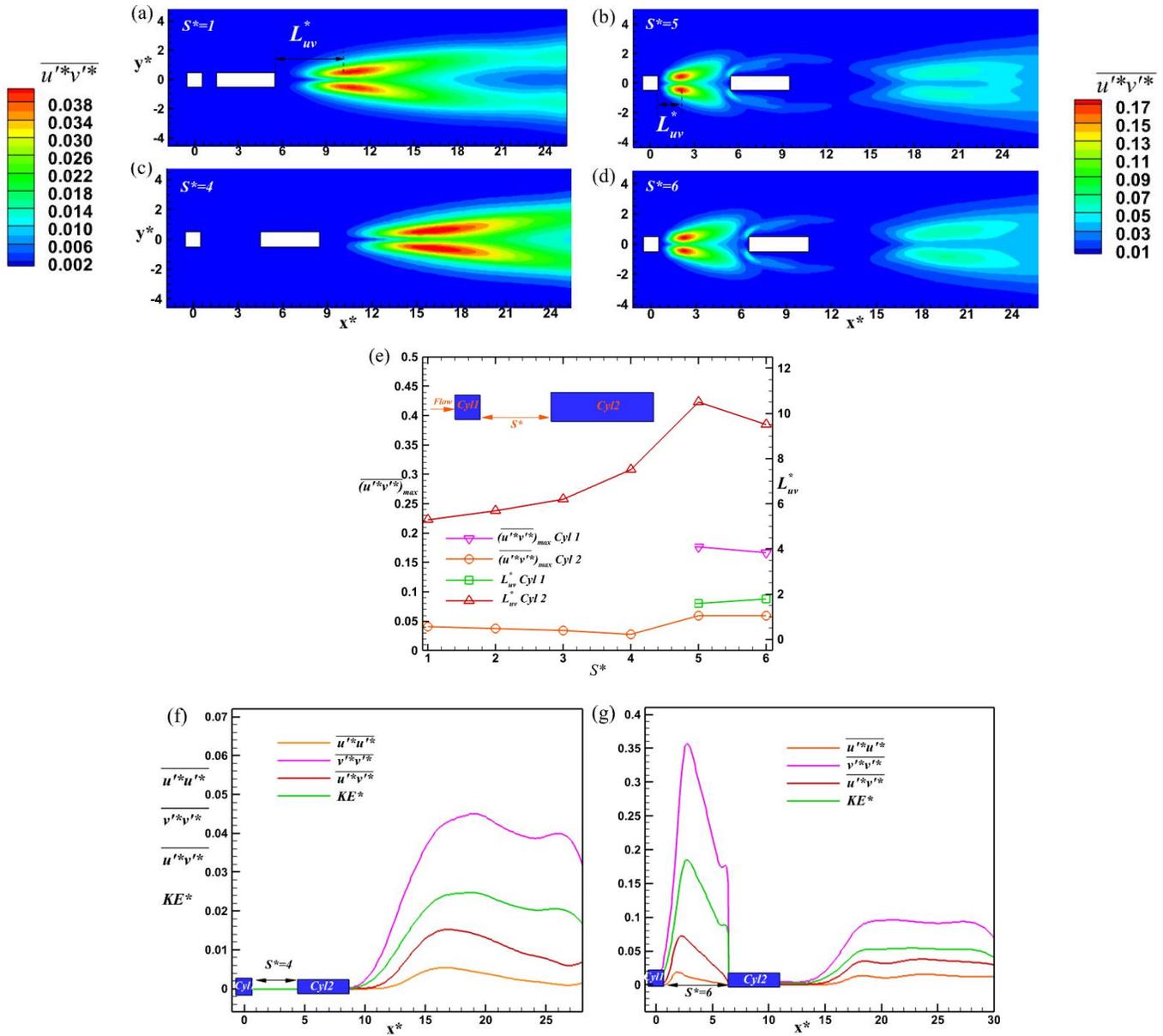

Figure 13. (a-d) The $\overline{u'^*v'^*}$ contours for various $S^*$, (e) the variations of $(\overline{u'^*v'^*})_{max}$ and $L^*_{uv}$ with $S^*$, the variations of $KE^*$, $(\overline{u'^*u'^*})$, $(\overline{v'^*v'^*})$ and $(\overline{u'^*v'^*})$ on the center line of the cylinders (y = 0) (f) $S^* = 4$, (g) $S^* = 6$, ($b_1^* = 1$, $b_2^* = 4$ and $Re = 150$).

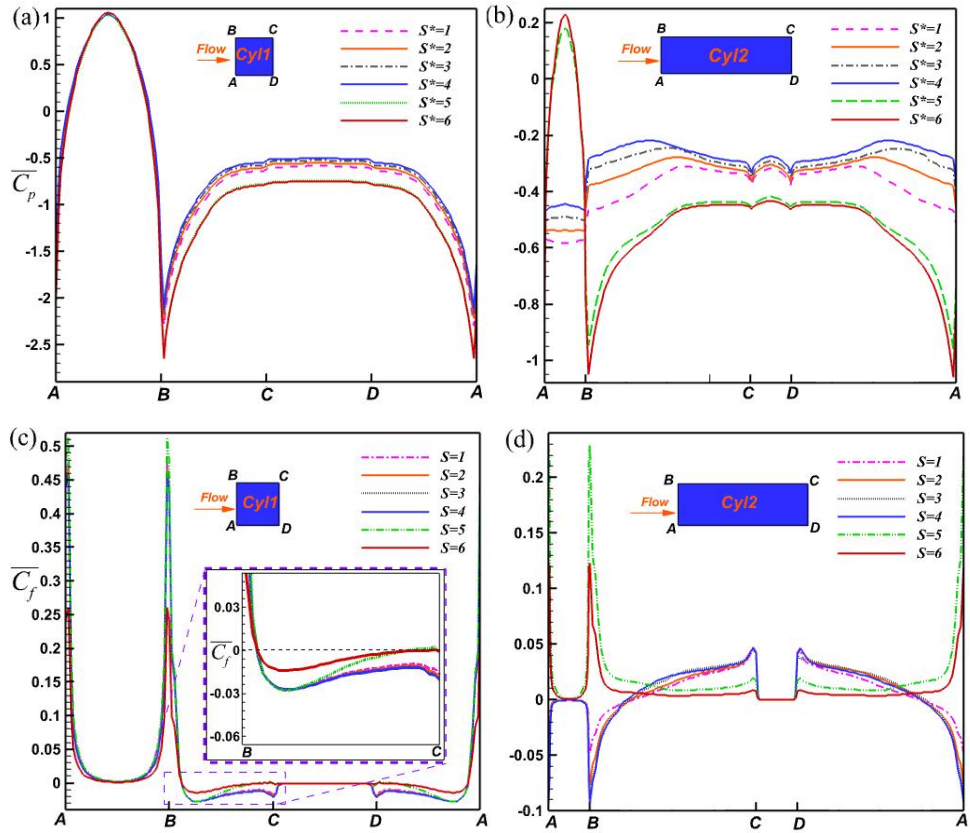

Figure 14. (a,b) Time-mean local pressure coefficient ($\overline{C}_p$) and (c,d) local friction coefficient ($\overline{C}_f$) for (a, c) upstream cylinder ($b_1^* = 1$) and (b, d) downstream cylinder ($b_2^* = 4$), $S^*$ =1-6 and $Re$ = 150.

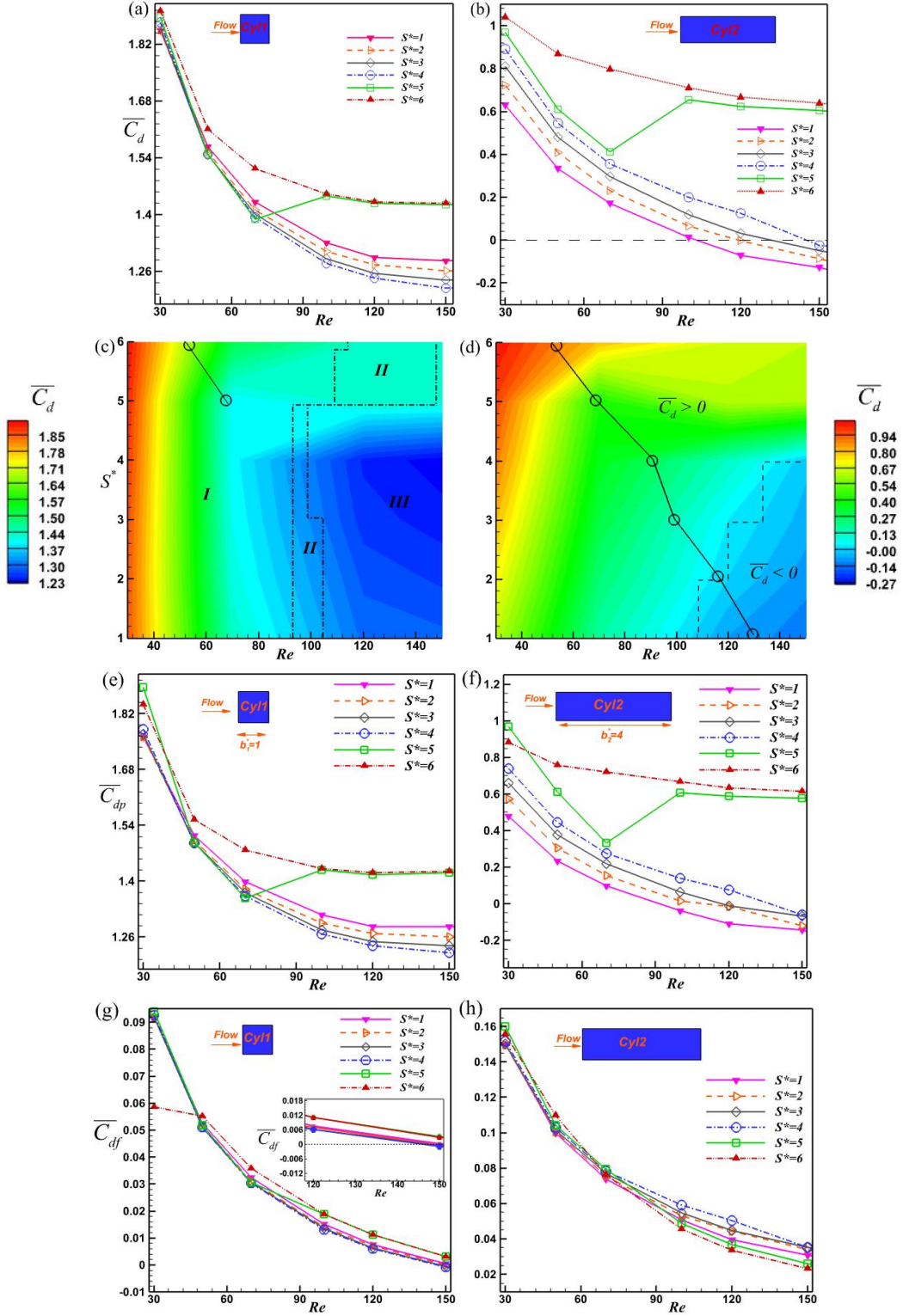

Figure 15. The profiles and contours of total drag coefficients, pressure drag coefficients profiles (e, f) friction drag coefficients profiles (g, h) for various $Re$ and $S^*$ and both cylinders. Upstream cylinder ($b_1^* = 1$, plots a, c, e, g) and downstream cylinder ($b_2^* = 4$, plots b, d, f, h). In the contours line with a circle, dashed lines, and long-dashed lines indicate $Re_{cr}$, the boundary between three mean flow patterns, and the boundary between positive and negative drags coefficient, respectively.

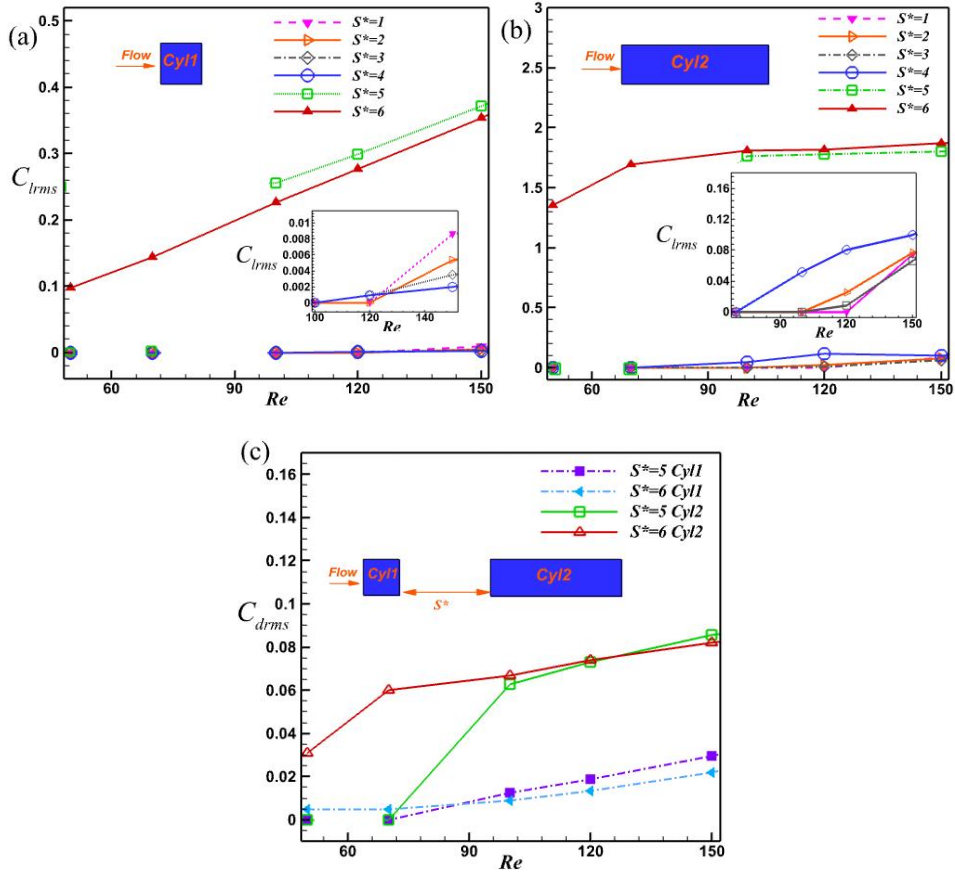

Figure 16. Variations of the $C_{lrms}$ and $C_{drms}$ for various $Re$ and $S^*$ (a) $C_{lrms}$ for upstream cylinder ($b_1^* = 1$), (b) $C_{lrms}$ for the downstream ($b_2^* = 4$) and (c) $C_{drms}$ for both cylinders.

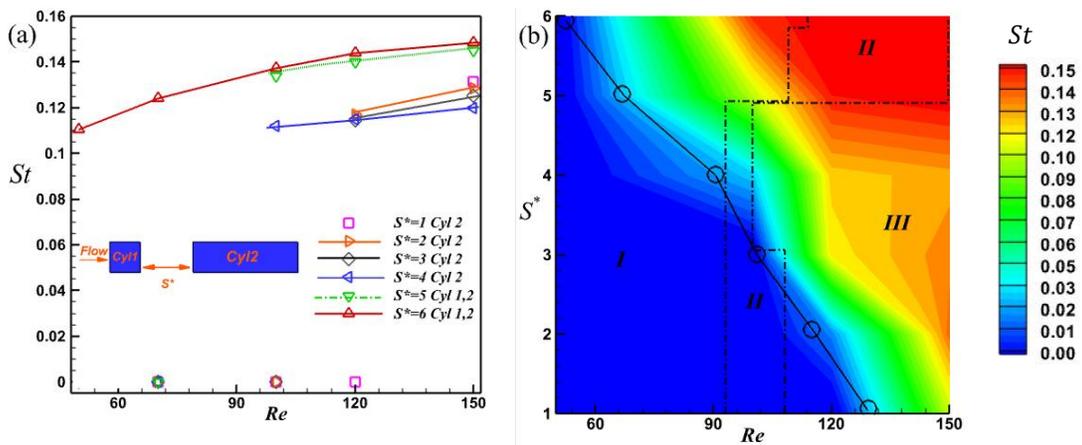

Figure 17. Variations of $St$ versus $Re$ and $S^*$ ($b_1^* = 1$, $b_2^* = 4$) in (a) profiles and (b) contour. The line with circle and dashed dot lines respectively indicate $Re_{cr}$ and the time-mean flow boundaries in the contour.

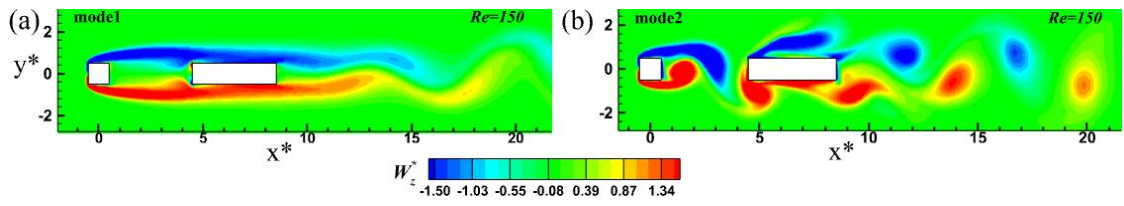

Figure 18. Vorticity contours for two tandem cylinders ($b_1^* = 1$, $b_2^* = 4$) at $Re = 150$ and $S^* = 4$ (a) Mode 1 (extended-body flow pattern or EBF), (b) Mode 2 (co-shedding flow pattern or CSF).

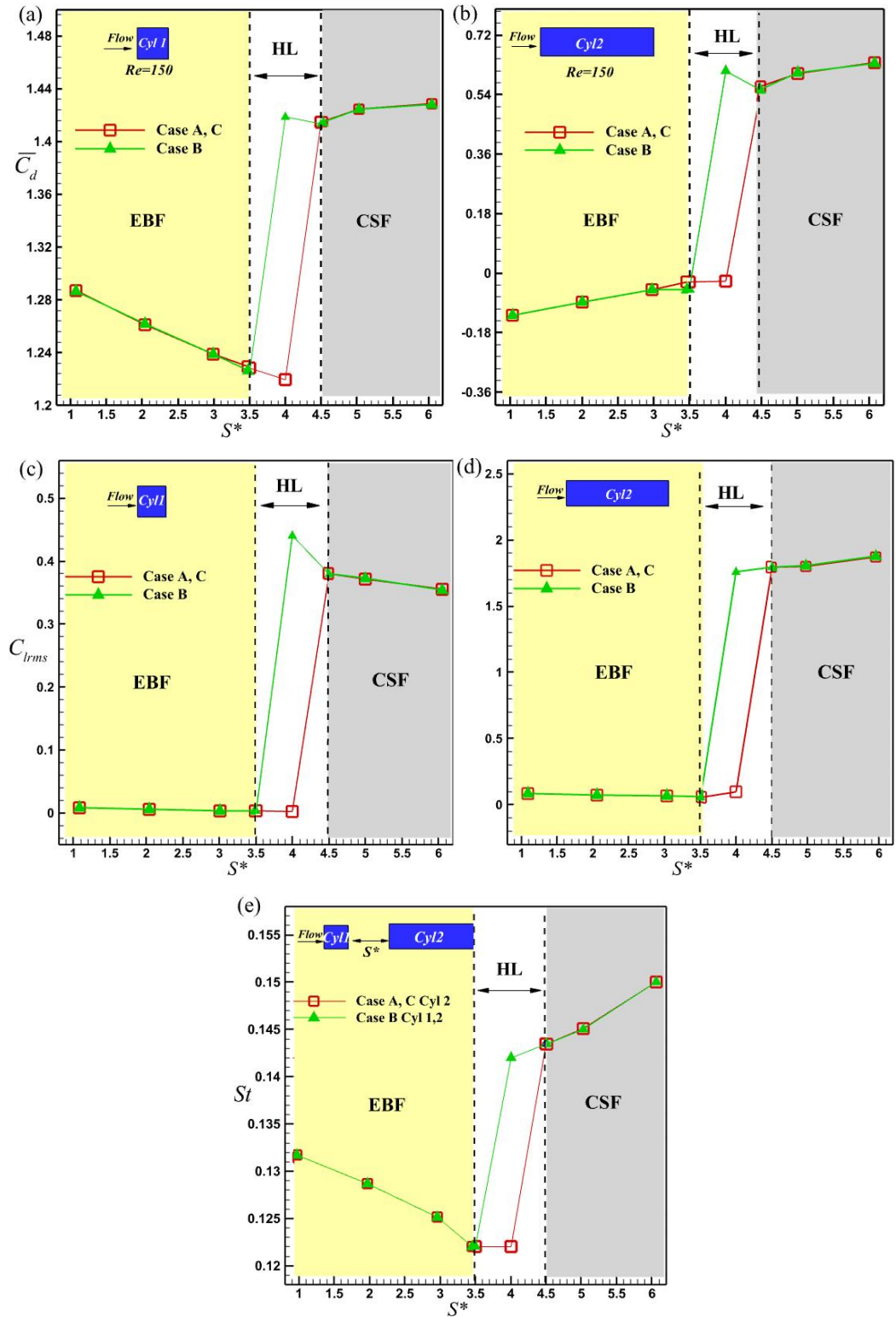

Figure 19. The variation of (a, b) $\overline{C}_d$, (c, d) $C_{lrms}$ and (e) $St$ ($b_1^* = 1$, $b_2^* = 4$) at $Re = 150$. EBF; extended-body flow pattern, HL; hysteresis limit, CSF; co-shedding flow pattern.

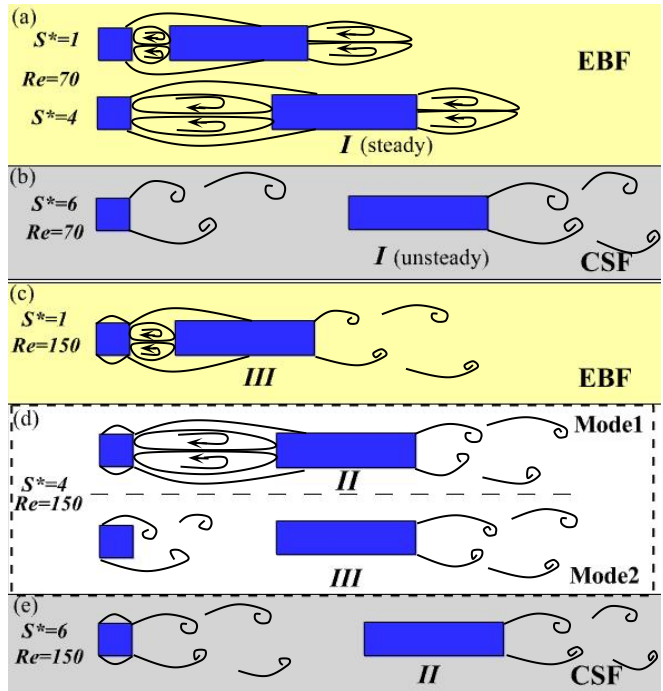

Figure 20. Overview of the flow patterns dependents on $S^*$ and $Re$ ($b_1^* = 1$, $b_2^* = 4$). EBF: extended-body flow pattern, CSF: co-shedding flow pattern.